\documentclass[iop]{emulateapj}
\usepackage[english]{babel}
\usepackage[latin1]{inputenc}
\usepackage{amsfonts}
\usepackage{graphicx}
\usepackage{color}

\shorttitle{The internal structure of overpressured RMHD jets}
\shortauthors{Mart\'{\i} et al.}

\begin{document}

\title{The internal structure of overpressured, magnetized, relativistic jets} 

\author{J. M. Mart\'{\i}\altaffilmark{1} and M. Perucho\altaffilmark{1}}
\affil{Departamento de Astronom\'{\i}a y Astrof\'{\i}sica, Universitat
de Val\`encia, 46100 Burjassot (Valencia), SPAIN}

\and

\author{J. L. G\'omez}
\affil{Instituto de Astrof\'{\i}sica de Andaluc\'{\i}a-CSIC, Glorieta
  de la Astronom\'{\i}a s/n, 18008 Granada, SPAIN}

\altaffiltext{1}{Observatori Astron\`omic, Universitat de
  Val\`encia, 46980 Paterna (Valencia), SPAIN}

\begin{abstract}
This work presents the first characterization of the internal
structure of
overpressured steady superfast magnetosonic relativistic 
jets in 
connection with their dominant type of energy.  To this aim, 
relativistic magnetohydrodynamic simulations of different jet models
threaded by a helical magnetic field have been analyzed covering a
wide region in the magnetosonic Mach number - specific internal energy
plane. The merit of this plane is 
that models dominated by different types of energy (internal energy:
hot jets; rest-mass energy: kinetically dominated jets; magnetic
energy: Poynting-flux dominated jets) occupy well separated
regions. The analyzed models also cover a wide range of
magnetizations. Models dominated by the internal
  energy (i.e., hot models, or Poynting-flux dominated jets with  
  magnetizations larger than but close to 1) have a rich internal
  structure {characterized by a series of recollimation shocks} and
  present the largest variations in the flow Lorentz 
  factor {(and internal energy density)}. Conversely, in
  {kinetically dominated} models there is 
  not much internal nor magnetic energy to be converted into kinetic
  one and the jets are featureless, with small variations in the flow
  Lorentz factor. The presence of a significant toroidal magnetic
  field threading the jet produces large gradients in the transversal
  profile of the internal energy density. {Poynting-flux dominated
    models with high magnetization} ($\approx 10$ or larger) are prone
  to be unstable against magnetic pinch modes, which sets limits to
  the expected magnetization in parsec-scale AGN jets {and/or
    constrains their magnetic field configuration}. 
\end{abstract}


\section{Introduction}

How relativistic jets are launched, accelerated, and collimated is
probably one of the most important questions related to AGN jet
physics and other astrophysical systems involving black hole
accretion, such as $\gamma$-ray bursts (GRBs) or tidal disruption
flares (TDFs). It is thought that dynamically important helical
magnetic fields twisted by the differential rotation of the black
hole's accretion disk or ergosphere play an important role
\citep{BZ77,BP82,MB09,TN11,ZC14}. As the jet propagates, part of the 
magnetic energy of the plasma is converted into kinetic energy,
accelerating the jet while maintaining a parabolic shape 
(see, e.g., \citealp{KB07}, and references therein for theoretical
approaches to the problem; see \citealp{NA13}, for an investigation of 
the parabolic jet structure in M87). For initially relativistic hot
jets, thermal acceleration can also play a role \citep[see, e.g.,][]
{GM95,GM97}. Simultaneous multi-wavelength and
Very Long Baseline Interferometry (VLBI) observations of AGN jets
suggest that the acceleration and collimation of the jet takes place
in the innermost $10^{4-6}$ Schwarzschild radii { from the central
  black hole}, upstream of the
millimeter VLBI (mm-VLBI) core \citep{MJ08}, defined as the bright
compact feature in the upstream end of the observed VLBI jet. 

  The simultaneity of multi-wavelength flares (from radio to
$\gamma$-ray energies) with the passage of a new superluminal
component through the mm-VLBI core has led to the suggestion  that
this corresponds to a strong recollimation shock
\citep[e.g.,][]{MJ08,MJ10,CG15a,CG15b}. {Moreover, in
  sources as CTA~102, in which this coincidence has not been proven,
  the presence of a stationary feature close to the VLBI core was
  claimed to explain the spectral evolution of a radio-flare
  \citep{FP11}. Multifrequency VLBI observations showed evidence in this
  direction \citep{FR13a,FR13b}}. The interaction of the moving
shock associated with the superluminal component and the standing 
shock at {or close to} the mm-VLBI core would produce
the particle acceleration and burst in particle and magnetic energy
densities required to produce the multi-wavelength flares. It should
be noted that this association of the mm-VLBI core with a
recollimation shock {would not be} in contradiction with the predictions from
the Blandford \& K\"{o}nigl jet model \citep{BK79} that establishes
the VLBI core as the location at which the jet becomes optically thin,
as long as this transition at {\it centimeter} wavelengths takes place
{\it downstream} of the mm-VLBI core. 
  
  Relativistic (magneto)hydrodynamical simulations have shown that
pressure mismatches between the jet and ambient medium lead to the
formation of a pattern of recollimation shocks
\citep[e.g.,][]{Wi87,DM88,DP93,GM95,GM97,GL16,MA09,PK15,MG15}. It is
therefore 
natural to expect that if the mm-VLBI core corresponds to a
recollimation shock other similar standing VLBI features would be
observed downstream of its location. Indeed, although some stationary
features have been found at hundreds of parsecs from the central
engine \citep[e.g.,][]{RG10}, most of the stationary components
observed in AGN jet appear in the innermost jet regions, close to the
VLBI core \citep[e.g.,][]{JM05,CM14,GL16}. Hence, obtaining a better
characterization of recollimation shocks is of special relevance not
only for the interpretation of the observed VLBI structure in AGN
jets, but also to obtain a better understanding of the nature of the
mm-VLBI core and its connection with the emission mechanisms at X and
$\gamma$-ray energies often observed from these sources.
    
  Recollimation shocks have been previously studied through
relativistic hydrodynamic and magnetohydrodynamic numerical
simulations
\citep[e.g.,][]{GM95,GM97,KF97,MM12,PK15,KP15,MG15,FP16}. In this 
paper we present the first systematic study of the resulting jet
structure in connection with the dominant type of energy in the jet, 
namely internal, kinetic, or magnetic, through relativistic
magnetohydrodynamical simulations of overpressured
superfast-magnetosonic jets { propagating through a homogeneous
  ambient medium. The effect of a pressure-decreasing atmosphere in
  the structure of jets, particularly in the properties of
  recollimation shocks, and the jet energy conversion will be the
  subject of future research}. {The paper is organized as follows. In
  Section~2, we define the parameter space of our study. Axisymmetric
  jet models are injected into the two-dimensional numerical grid in
  transversal equilibrium to minimize radial perturbations. In
  Section~3 we describe the transversal structure of the injected
  models. Section~4 is devoted to describe the setup of the
  simulations, whereas in Section~5 we present and discuss the results
  on the internal structure of jets. Finally, in Section~6 we
  summarize our main conclusions.}  
 
\section{Parameter space}
\label{s:param}

  In the purely hydrodynamical, Newtonian case, the basic
parameters governing the propagation of a supersonic, initially
cylindrical jet with purely axial speed across an homogeneous ambient
medium at rest can be taken as \citep[see, e.g.,][]{NW82} the jet
density, $\rho_j$, the jet overpressure factor, $K$, and the internal
jet Mach number, $M_j$. Models are expressed in units of the ambient medium
density and pressure, $\rho_a$, $p_a$, and the jet radius at
injection, $R_j$. 

  In the relativistic case, the presence of the light speed, $c$, as a 
constant appearing in the hydrodynamical equations, spawns an 
additional parameter and the simulations are defined by means of
\citep[see, e.g.,][]{MM97} $\rho_j$, $K$, the (axial) velocity of
the flow in the jet, $v_j$, and the classical or relativistic internal
jet Mach number, $M_j$, ${\cal M}_j$, respectively, in units of the
ambient density and the jet radius at injection.

  In the RMHD case, assuming that the radial magnetic field is zero at 
injection, two new quantities are needed to define the magnetic field 
configuration, namely the azimuthal and axial magnetic field
components, $B^\phi_j$, $B^z_j$, or equivalently, the jet
magnetization, $\beta_j$, and the magnetic pitch angle, $\phi_j$. 
{The jet magnetization is defined as $\beta_j =
  p_{m,j}/p_j$, where $p_{m,j}$ and $p_j$ stand, respectively, for the
  jet magnetic pressure and the jet thermal (or gas) pressure. The
  magnetic pressure is defined as $p_{m,j} = b^2_j/2$, where $b^2_j$
  is the magnetic energy density\footnote{{Quantity $b^2$ stands for $b^\mu
    b_\mu$, where $b^\mu$ ($\mu = t, r, \phi, z$) are the components
    of the 4-vector representing the magnetic field in the fluid rest
    frame, and summation over repeated indices is assumed.}}.} On
the other hand, in the case of supermagnetosonic jets as those
considered here, the role of the Mach number will be played by the
magnetosonic Mach number, ${\cal M}_{ms,j}$ (see the Appendix). 
Together with other parameters (significantly the jet overpressure
factor, $K$), the relativistic magnetosonic Mach number governs the
properties of internal conical shocks in overpressured
magnetized jets in the same way as the Mach number does in purely
hydrodynamic, overpressured jets. In this work, units are used in which the
light speed, the ambient density and the jet radius at injection are
set to unity. Besides that, a factor of $\sqrt{4 \pi}$ is absorbed in
the definition of the magnetic field. Finally, both the jet and the
ambient medium plasmas are assumed to behave as a perfect gas with
constant adiabatic index, $\gamma = 4/3$. {This value
  (which corresponds to the adiabatic index in the ultrarelativistic
  limit) is inappropriate to describe thermodynamically the plasma in
  the cold models. However, in these cases, the internal energy is
  negligible and to overestimate it by a factor of two with respect to
  the non-relativistic value obtained with an adiabatic index of $5/3$
  has no qualitative effects on the jet dynamics. The adiabatic index
  of $4/3$ is also inadequate to describe the ambient medium but in
  these simulations where the ambient medium is static and fixed to 
  its initial values, the effect of using one adiabatic index or
  another can be absorbed in the definition of plasma density.} 

%
\begin{table*}
 \centering
 \begin{minipage}{145mm}
\small
  \caption{Parameters defining the overpressured jet models.}
  \begin{tabular}{lcccrcc|c|ccc}
  \hline
\\
  Model & $\rho_j \, [\rho_a]$ & $K$& $v_j \, [c]$ & ${\cal M}_{ms,j}$ & $\beta_j$ &
 $\phi_j$ [$^\circ$] & $\Delta r_{sl}^a \,[R_j]$ & $\varepsilon_j \, [c^2]$
 & $K_1^b$ & $p_a \, [\rho_a c^2]$ \\
\\
  \hline \\
  PH02 & $5 \times 10^{-3}$ & $2$ & $0.95$ & $2.0\phantom{00}$ & $2.77\phantom{0}$ &
  $45.0$ & 0.12 & 10.0\phantom{0000} & $1.87$ & $3.31 \times 10^{-2}$\\
  PK02 & $5 \times 10^{-3}$ & $2$ & $0.95$ & $2.0\phantom{00}$ & $10.0\phantom{000}$ &
  $45.0$ & 0.49 & 0.458\phantom{0} & $1.84$ & $4.20 \times 10^{-3}$\\ 
  HP03 & $5 \times 10^{-3}$ & $2$ & $0.95$ & $3.5\phantom{00}$ & $0.454$ &
  $45.0$ & 0.12 & 10.0\phantom{0000} & $1.94$ & $1.21 \times 10^{-2}$\\ 
  PK03 & $5 \times 10^{-3}$ & $2$ & $0.95$ & $3.5\phantom{00}$ & $10.0\phantom{000}$ &
  $45.0$ & 0.49 & 0.117\phantom{0} & $1.84$ & $1.08 \times 10^{-3}$\\ 
  KH06 & $5 \times 10^{-3}$ & $2$ & $0.95$ & $6.0\phantom{00}$ & $0.5\phantom{00}$ &
  $45.0$ & 0.49& 0.468\phantom{0}  & $1.94$ & $5.85 \times 10^{-4}$\\ 
  KP06 & $5 \times 10^{-3}$ & $2$ & $0.95$ & $6.0\phantom{00}$ & $10.0\phantom{000}$ &
  $45.0$ & 0.32 & 0.0317 & $1.84$ & $2.90 \times 10^{-4}$\\ 
  KH10 & $5 \times 10^{-3}$ & $2$ & $0.95$ & $10.0\phantom{00}$ & $0.5\phantom{00}$ &
  $45.0$ & 0.24  & 0.133\phantom{0} & $1.94$ & $1.66 \times 10^{-4}$\\ 
  KP10 & $5 \times 10^{-3}$ & $2$ & $0.95$ & $10.0\phantom{00}$ & $10.0\phantom{000}$ &
  $45.0$ & 0.24 & 0.0132 & $1.84$ & $1.21 \times 10^{-4}$\\ 
\\
\hline
\end{tabular}
\label{t:t1}
$^a$ $\Delta r_{sl}$ is the width of the shear layer defined as the
radial section of the jet where the function defining the transition,
sech$(r^m)$, is between 0.1 and 0.9, corresponding to jet mass
fractions between 0.1 and 0.9.\\
$^b$ $K_1$ stands for the total (gas plus magnetic) overpressure
factor at the jet surface (see text). 
\vspace{2mm}
\end{minipage}
\end{table*}
%

  Table~\ref{t:t1} displays the values of the six parameters (namely
$\rho_j$, $v_j$, $K$, ${\cal M}_{ms,j}$, $\beta_j$ and $\phi_j$)
defining the models.  Given the type of
transversal equilibrium profiles considered in this work (obtained for
specific profiles of the azimuthal magnetic field as discussed in the
next section), $K$, ${\cal M}_{ms,j}$, $\beta_j$ and $\phi_j$
represent jet cross section averages. All the jet models have the same
rest-mass density and flow velocity, and the same average magnetic
pitch angle and overpressure factor. On the contrary, the relativistic
magnetosonic Mach number changes by a factor of 5 and the
magnetization, by a factor of 20, among the different jet
models. {Note that all the models have initial
  toroidal speeds equal to zero and, consequently, are better suited
  to describe the jets at distances far beyond the jet formation region.}
Table~\ref{t:t1} also displays some derived parameters, as the
pressure mismatch at the jet surface, $K_1$, the ambient pressure,
$p_a$, and the specific internal energy in the jet,
$\varepsilon_j$. The ambient pressure changes more than two orders of 
magnitude, although its value is always small compared with the
rest-mass energy density of the ambient medium. The values of the
specific internal energy in the jet span three orders of magnitude
including cold as well as hot jet models. Finally, the transition
between the jet and the ambient medium is smoothed by means of a shear
layer of width $\Delta r_{sl}$ by convolving the sharp jumps with the
function sech$(r^m)$, where $m \in [4,16]$. Different widths of the
shear layer are needed to stabilize the models against pinch
instabilities (see Sect.~\ref{ss:sl}).

%
\begin{figure*}
\begin{minipage}{180mm}
\begin{center}
\includegraphics[width=8.6cm,angle=90]{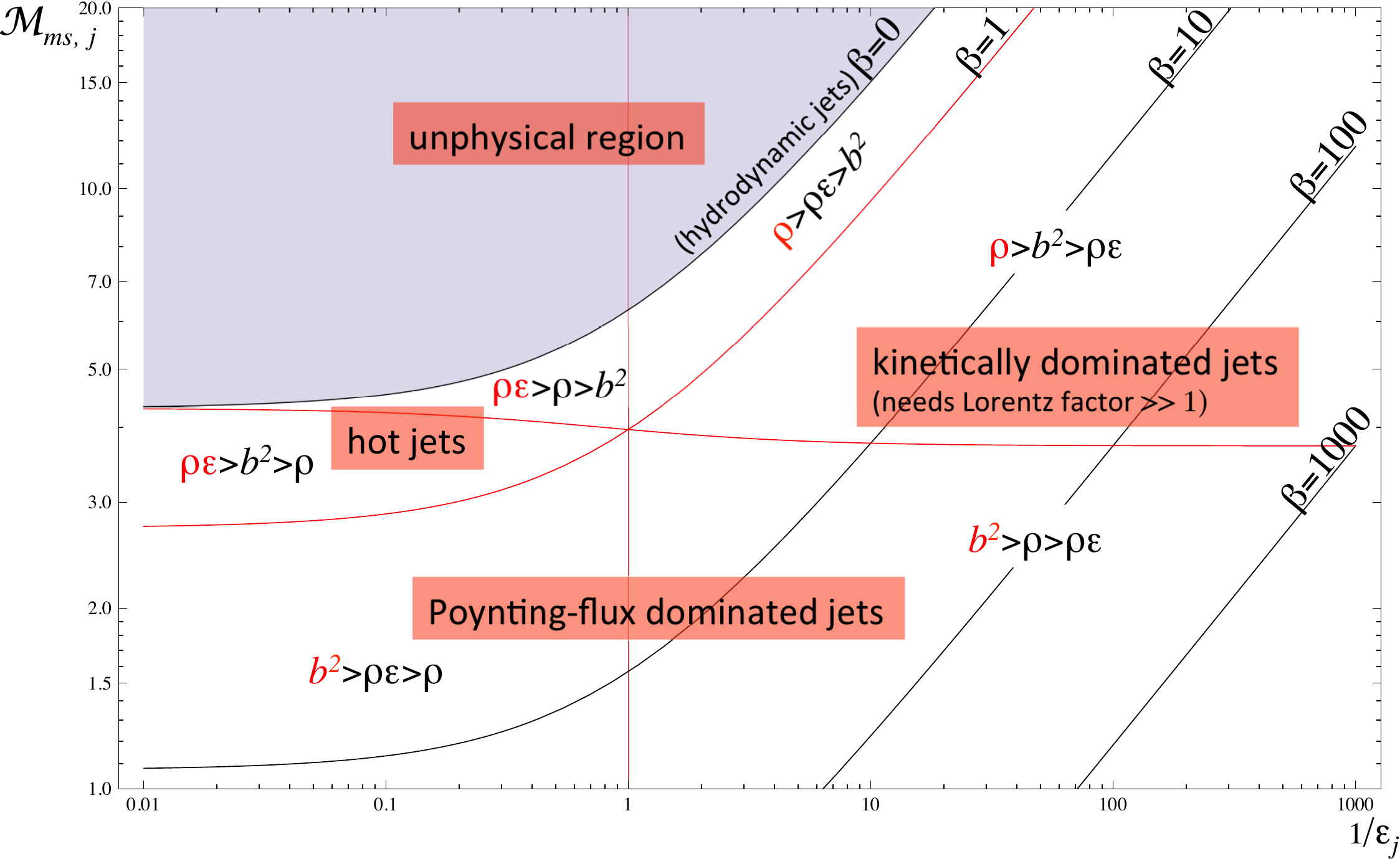} 
\includegraphics[width=8.6cm,angle=90]{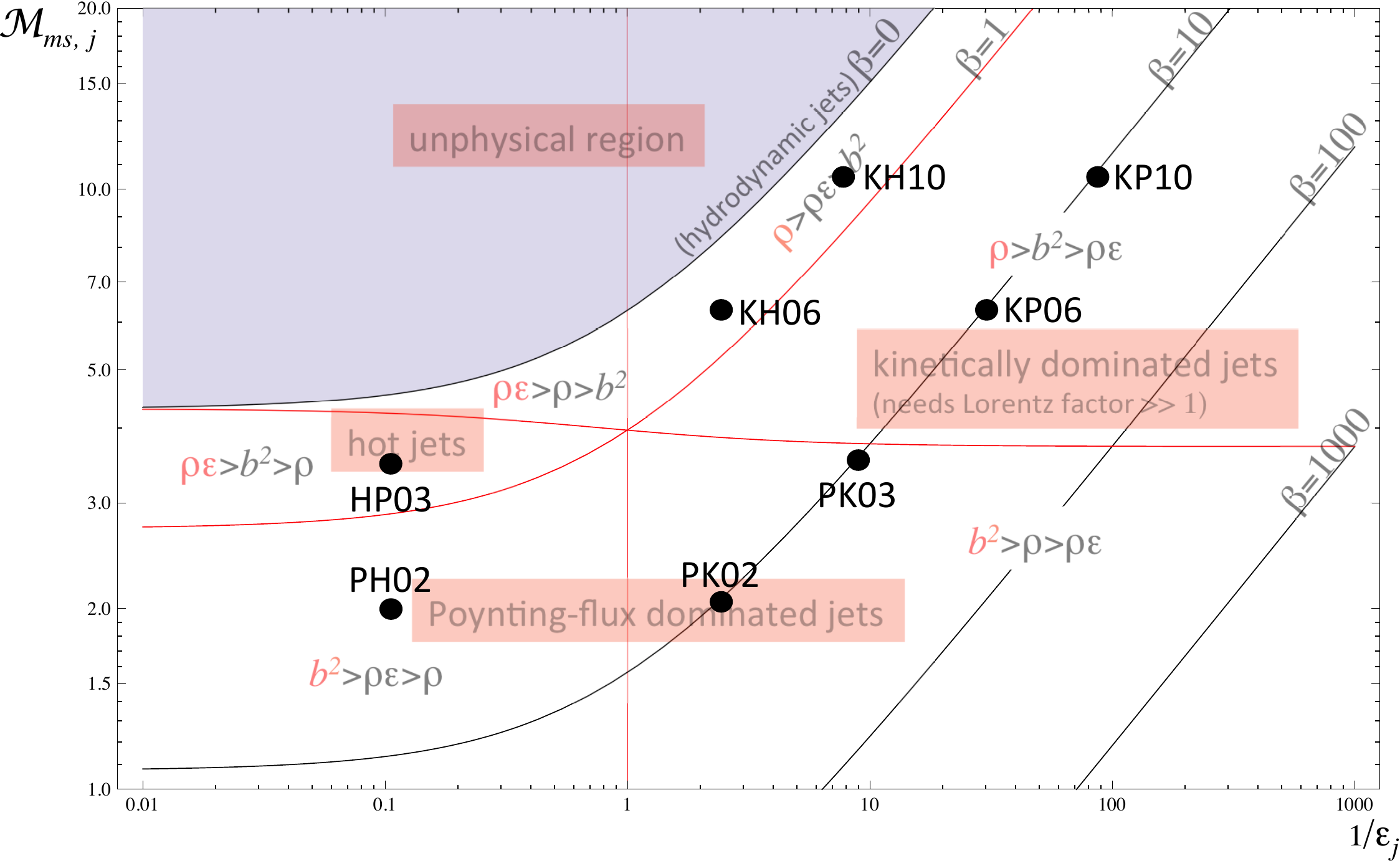}
\caption{Top panel: Jet classes in the ${\cal M}_{ms,j} - 1/\varepsilon_j$ plane
  according to the dominant energy type, 
  for given jet flow velocity, $v_j= 0.95$. Drawn are lines of
constant magnetization (0, 1, 10, 100, 1000). Kinetically dominated
jets, Poynting flux dominated jets and hot jets are placed in the
diagram with the help of three (red) lines corresponding to models
with $\rho_j = \rho_j \varepsilon_j$, $b_j^2 = \rho_j \varepsilon_j$,
$b_j^2 = \rho_j$. Pure hydrodynamic models are placed on the $\beta_j
= 0$ line which bounds a forbidden region (in violet) corresponding to
unphysical models with negative magnetic energies. Bottom panel:
Distribution on the ${\cal
  M}_{ms,j}-1/\varepsilon_j$ diagram of the models considered in this paper.}
\label{f:f1}
\end{center}
\end{minipage}
\end{figure*}
%

  The parameters of the models are chosen to span a wide region in the ${\cal
M}_{ms,j}$-$1/\varepsilon_j$ plane (see Fig.~\ref{f:f1}). According to
the type of energy flux that dominates, jet models 
can be classified as kinetically dominated (those models dominated by
the rest-mass energy, $\rho_j>\max(\rho_j \varepsilon_j, b_j^2)$ and
Lorentz factor $W_j \gg 1$), internal energy dominated (or hot jets,
$\rho_j \varepsilon_j >\max(\rho_j, b_j^2)$) and Poynting flux
dominated ($b_j^2>\max(\rho_j, \rho_j \varepsilon_j)$). The plane
displayed in Fig.~\ref{f:f1} has the virtue of placing these three
types of models in well separated regions\footnote{ The division
  between the regions is based on the averaged values 
of the radial profiles of the quantities defining the jet models and
it can certainly be understood as {\it universal}
for this set of variables, i.e., the lines separating the different 
energy regimes rely on a series of analytic expressions for (the averaged 
values of) the variables defining the jet and ambient media. Of course the 
particular diagram does depend on the functional dependence chosen for 
the radial profiles, and a number of free parameters (like the adiabatic 
index of the equation of state, the jet overpressure factor, the jet flow 
velocity, the jet-to-ambient rest-mass density ratio and the magnetic pitch angle). 
Finally, it also depends on a simplified definition (direction-independent) of the magnetosonic 
speed (see the Appendix).}. 
Our current understanding of the process of jet
acceleration indicate that jets would form at
some point in the hot/Poynting-flux dominated region and would evolve
towards the region of kinematically dominated models \citep[see, e.g.,
][and references therein]{KB07}. When projected 
onto this diagram, the simulations discussed by \cite{RP08,RP09} would
be placed on top of the line $1/\varepsilon_j = 1$ with magnetizations
$\beta_j \in [0.1, 10]$, whereas those of \cite{MG15} will be on the
line $1/\varepsilon_j = 0.095$ with magnetizations $\beta_j \in
[0,0.4]$. In all the cases, the models are in the hot models region or
its neighbourhood. 

  Names are given to the models according to the following rule: two capital
letters to indicate the two dominating energy types (``K'', for
kinetically dominated jets; ``P'', for
Poynting-flux dominated jets; ``H'', for hot jets) in order of
prevalence, and two digits
related with the Mach number of the jet flow.

\section{Transversal structure of the injected jet models}
\label{s:transversal}

Jets are injected in internal transversal equilibrium to minimize the
sideways perturbations once immersed in the ambient medium and to
obtain an internal structure as clean as possible. The profiles of the
rest-mass density, the axial flow velocity and the axial magnetic
field across the jet are taken constant.

  The azimuthal magnetic field in the laboratory frame is defined
according to
\begin{equation}
B^\phi(r) = \left\{ \begin{array}{ll}
\displaystyle{\frac{2 B_{j, \rm
    m}^\phi (r/R_{B^\phi, \rm m})}{1 + (r/R_{B^\phi, \rm
    m})^{2}}}, & 0 \leq r \leq 1 \\

0,        & r > 1.
               \end{array} \right.
\label{eq:bphi}
\end{equation}
This function represents a toroidal magnetic field that grows
linearly for $r \ll R_{B^\phi, \rm
    m}$, reaches a maximum ($ B_{j, \rm
    m}^\phi$) at $r = R_{B^\phi, \rm
    m}$, then decreases as $1/r$ for $r \gg R_{B^\phi, \rm
    m}$ and is set equal to zero for $r>1$. It is a smooth fit of the
  piecewise profile used by \cite{LP89} \citep[see also][]{Ko99a, LA05}
  and corresponds to a uniform current density for radius $r \ll
  R_{B^\phi, \rm m}$, declining up to $r = 1$, and a return
  current at the jet surface. { The radius at which the toroidal
    magnetic field reaches its maximum, $R_{B^\phi, \rm
    m}$, has been fixed to 0.37 in all the models.}

  In the case of a jet without rotation, the
equilibrium equation for the transversal equilibrium can be written
\citep[e.g.,][]{Ma15} 
\begin{equation}
\frac{d p}{d r} = -\frac{(B^\phi)^2}{r
    W^2} - \frac{B^\phi}{W^2} \frac{d B^\phi}{d r},
\label{e:peq}
\end{equation}
where $p$ is the gas pressure and $W$, the jet Lorentz
factor (corresponding in this case to a purely axial flow). This
equation can be integrated by separation of variables to give 
\begin{equation}
p(r) = \left\{ \begin{array}{ll}
\displaystyle{2 \left(\frac{B_{j, \rm
    m}^\phi}{W(1 + (r/R_{B^\phi, \rm
    m})^{2})}\right)^2 + C}, & 0 \le r \le 1 \\

p_a', & r > 1,
               \end{array} \right.
\end{equation}
where we choose $p_a' = K p_a$ (with $K>1$) to obtain equilibrium models of
overpressured jets. Using the boundary condition $p^*_1 = p_a'$,
where $p^*$ is the total (gas plus magnetic) pressure and $p^*_1$
stands for $p^*(r \! = \! 1)$, the integration constant $C$ can be
fixed to be
\begin{equation}
C = p_a' - \frac{(B_j^z)^2}{2} - \frac{(B^\phi_1)^2}{2W^2}
(1 + (R_{B^\phi, \rm m})^2).
\end{equation}

\begin{figure*}
\begin{minipage}{176mm}
\begin{center}
\includegraphics[width=14.6cm,angle=0]{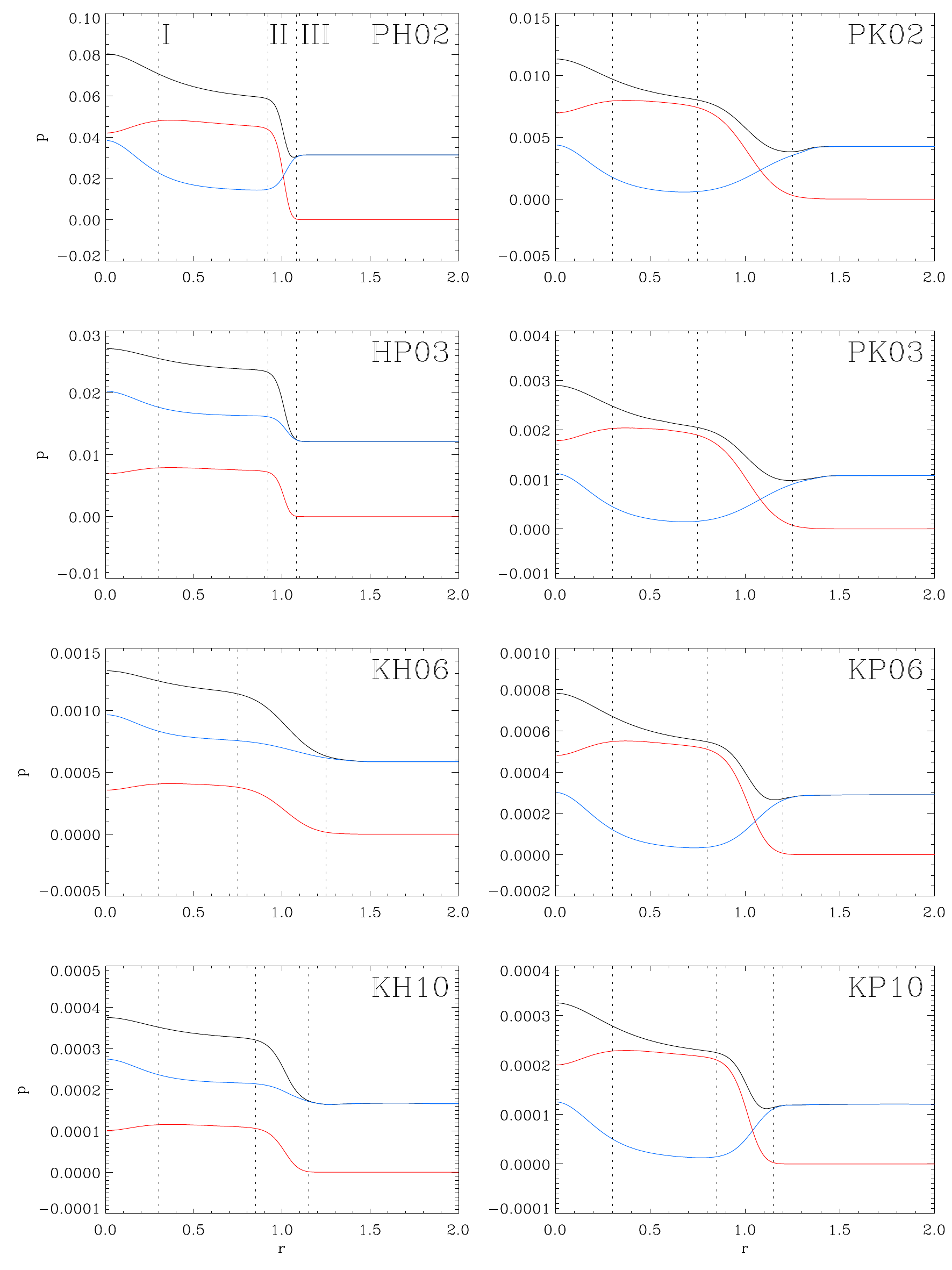}
\caption{Gas (blue filled line), magnetic (red filled line) and total pressure
  (black filled line) across the jet at injection for the eight models
  considered in this work. Models with the same magnetization (HP03,
  KH06 and KH10, on one side; PK02, PK03, KP06 and KP10 on the other)
  have the same pressure profiles (with slight variations due to the
  different widths of the shear layers). In models with
magnetization smaller than $1.0$ (HP03, KH06, KH10), gas pressure
dominates over magnetic pressure across the jet. Vertical dotted lines
define the layers making up the transversal jet structure: $0<r<r_{\rm
  I}$: hot central spine; $ r_{\rm II} < r < r_{\rm III}$: shear layer.}
\label{f:pressure}
\end{center}
\end{minipage}
\end{figure*}

  This transversal structure is convolved with a shear layer to smooth
the transition between the jet and the ambient (see
Sect.~\ref{s:param}). { It is interesting to note that by
  introduzing this shear layer, the current sheet at the jet surface is removed.} 
Fig.~\ref{f:pressure} shows the (gas, magnetic and total) pressure 
profiles across the jets (including the shear layers) for the models
considered in this paper. For our choice of parameters, the magnetic pressure 
\begin{equation}
p_m(r) = \frac{{B_j^\phi(r)}^2}{W_j^2} + (B_j^z)^2,
\end{equation}
is dominated by the (constant) contribution of the axial component of
the magnetic field, although it is modulated by the profile of the
toroidal component. Hence, it has a minimum at the jet axis, and then
increases up to $r=R_{B^\phi, \rm m}$ where it has the maximum. Beyond
this point, the magnetic pressure decreases slowly with radius up to
the surface.  

  Contrary to the magnetic pressure, the gas pressure has a local
maximum at the jet axis and decreases progressively
faster up to $r=R_{B^\phi, \rm m}$. At larger radii, 
and before entering into the shear layer, the gas pressure tends to a
constant value\footnote{For $r>R_{B^\phi, \rm m}$, $B^\phi$  decreases
  aproximately as $1/r$, which, according to Eq.~\ref{e:peq}, leads to
  a constant value of the gas pressure.}. Both thermal and magnetic
pressure combine to produce the monotonically decreasing radial
profile in the total pressure needed to balance the magnetic tension.

Although based on a particular choice of parameter
  profiles, the general conclusion is that the existence of a magnetic
  field with a significant toroidal component produces a complex
  transversal structure in magnetized jets with a central spine
  (extending up to the radius where the maximum of the magnetic 
  tension is reached, some point between $r=0$ and $r=R_{B^\phi, \rm
    m}$) where the thermal pressure (and hence the plasma internal
  energy) is close to its maximum. {A layer with milder (magnetic,
    thermal) pressure profiles surrounds this central spine. This
    layer extends up to the outer jet/ambient-medium shear layer (see
    Fig.~\ref{f:pressure}).} 
 
  Finally, it can be easily seen that the presence of a toroidal field like the
one defined in (\ref{eq:bphi}) increases the gas pressure up to $r =
\sqrt{R_{B^\phi, \rm m} ((R_{B^\phi, \rm m})^2 - 
  R_{B^\phi, \rm m} +1)}$ ($\approx 0.53$, for $R_{B^\phi, \rm m} =
0.37$) and decreases it outside so that the average gas pressure inside
the jet remains unchanged with respect to the case of zero toroidal
magnetic field \citep{Ma15}.

%
%

\section{Numerical simulations}

The numerical RMHD code used in these simulations is a second-order,
conservative, finite-volume code based on high-resolution
shock-capturing techniques. An overview of the specific algorithms
used in the code and an analysis of its performance can be found in
{ Appendices} A and B, respectively, of \cite{Ma15}.

  The equilibrium profiles discussed in the previous section are
used as a boundary condition to inject the jets into a two-dimensional
domain representing an ambient medium with a pressure mismatch.
In their attempt to reach again the equilibrium, the jets undergo
sideways motions generating radial components of the flow velocity
and the magnetic field that break the slab symmetry of the original
jet model along the $z$-axis.

  The jets are injected through a nozzle of radius $R_j$ equal to 1
into an axisymmetric cylindrical domain with $(r,z) \in [0, L_r]
\times [0, L_z]$, with $L_r =6$ and $L_z = 60, 80, 120$, depending on
the spacing of the shocks in each model. The evolution of the flow
in the domain is simulated with the RMHD code in 2D radial, 
axial cylindrical coordinates with a resolution of 80 (40) cells per
jet radius in the radial (axial) direction. In order to disturb the
ambient medium as little as possible along the simulation, the domain 
$(r,z) \in [0, 1] \times [0, L_z]$ is initially filled with the
analytical, injection solution. Reflecting boundary conditions are set
along the axis ($r = 0$, $z>0$) and at the jet base outside the
injection nozzle ($r >1$, $z = 0$). Zero gradient conditions are set
in the remaining boundaries. 

 The models, set up to be in equilibrium with an ambient pressure
$p'_a = K p_a$ ($K>1$), are injected into an atmosphere with pressure
$p_a$. The new equilibrium states are set through a series of conical
fast-magnetosonic shocks which are the subject of study of the present
paper. { Reaching such an equilibrium state is a lengthy process
  that typically takes between 3 and 5 axial grid light crossing times
  ($2-6 \times 10^4$ time iterations) per simulation. In computational
  time this means about 10 to 50 days of single-processor CPU time. This time
  was reduced in practice by a factor of 10 using an OpenMP parallel
  version of the code with 12 processors. During this transient phase
  the flow  suffers (more or less violent) axially-symmetric sideways
  expansions and compressions which in some cases had to be damped out
  with the help of the shear layer to avoid the growth of pinch
  instabilities (see Sect.~\ref{ss:sl}).} 

\section{Results}

\subsection{Overall jet structure}

   The steady state jets corresponding to the models defined in
Table~\ref{t:t1} are shown in Figs.~\ref{f:PH02} to \ref{f:KP10}. Each
figure contains panels displaying the distributions of rest-mass
energy density and gas pressure (both in logarithmic scale), flow
Lorentz factor (with the poloidal streamlines overimposed), and 
toroidal and axial magnetic field components (with the poloidal
magnetic field lines overimposed on the axial magnetic field
panel). Besides these maps another one displaying 
the toroidal flow speed generated during the jet evolution is also
shown. Small radial components of the flow speed and the magnetic
field are also generated during the flow evolution but are not shown,
although their magnitude relative to that of the corresponding axial
component can be inferred from the bending of the poloidal lines. Some
general conclusions can be extracted from the analysis of these
figures: 

\begin{enumerate}
\item In all the models, the equilibrium of the jet is established
  by a series of expansions and compressions of the jet flow against
  the ambient medium. Standing oblique shocks { (recollimation
    shocks)} associated with these 
  jet oscillations can be distinguished in some cases, specially in 
  hot models (PH02, HP03) and, to a lesser extent, in colder, low
  magnetization models (KH06, KH10). A more quantitative analysis of
  the jet oscillations and the standing shocks is presented in
  Sects.~\ref{ss:sl} and~\ref{ss:is}, respectively.

\item As a consequence of the profile of the magnetic pressure across
  the jet and, specially, of the magnetic pinch exerted by the
  toroidal magnetic field, the thermal pressure is not constant across
  the jet (see Fig.~\ref{f:pressure} and the accompanying discussion on the
  transversal structure of the injected jet models in
  Sect.~\ref{s:transversal}). Models with large magnetizations (PK02, 
  PK03, KP06, KP10) concentrate most of their internal energy in a
  thin hot spine around the axis (see the panels of gas
  pressure in Figs.~\ref{f:PK02}, \ref{f:PK03}, \ref{f:KP06} and
  \ref{f:KP10}), as discussed in Sect.~\ref{s:transversal}. 

\item Despite the large difference in magnetization (a factor of
  20), kinetically dominated jet models  KH10 and KP10 have
  very similar overall structure (jet oscillation, amplitude of
  variations, local jet opening angles,...) exception made of the
  already mentioned central hot (in relative terms) spine in the KP10
  jet. In these kinetically dominated models there is no significant
  internal nor magnetic energy to convert into kinetic, and the flow
  Lorentz factor is virtually constant despite the wide jet sideways
  oscillations. 

\item All the models develop small azimuthal velocities (of the
  order of 2\% of the speed of light or smaller). These speeds tend to
  be larger in those models with larger maximum local opening angles
  (again hot models and neighbours).

The Lorentz force acting on the relativistic magnetized fluid is 
\begin{equation}
{\bf F}_L = {\bf J} \times {\bf B} + \rho_e {\bf E},
\end{equation}
where ${\bf J} = \nabla \times {\bf B}$ and $\rho_e = \nabla \cdot
{\bf E}$ stand, respectively, for the current and
electric charge densities, and ${\bf E} = - {\bf v} \times {\bf B}$ is
the electric field (in the ideal MHD approximation). In cylindrical
coordinates $(r, \phi,z)$, and for an axisymmetric flow, the azimuthal
component of the Lorentz force can be worked out to be
\begin{eqnarray}
\nonumber
F_L^\phi  & =  B^z \displaystyle{\frac{\partial B^\phi}{\partial z}}  & + \frac{B^r}{r}
\frac{\partial (rB^\phi)}{\partial r} \\
& & + \rho_e (v^r B^z + v^z B^r).
\end{eqnarray}

Despite the fact that the considered jet models are injected into the
numerical domain with purely axial flow velocities, the development of
a radial component of the velocity and the magnetic field, and an
axial dependence of the toroidal magnetic field, as a result of the
transversal equilibrium mismatch between the injected jet and the ambient
medium, {produce} a net toroidal force that causes the growth of
non-zero toroidal flow speeds.  

\end{enumerate}

\subsection{Effects of the shear layer and detailed jet structure}
\label{ss:sl}

  Before setting into their final steady solutions, the overpressured jet
models undergo a transient phase in which the flow suffers (more or
less violent) axially-symmetric sideways expansions and
compressions. In some cases, remarkably those corresponding to cold, 
Poynting flux dominated jet models (PK02, PK03) and to a less extent
kinetically dominated jet models (KH06, KP06\footnote{The same happens
  to kinetically dominated models KH10, KP10 but for larger axial
  distances.}), the pinch exerted at some points of the jet axis
during this transient phase \citep[{due to the
  coupling of the sideways oscillation caused by the jet overpressure
  with current driven instabilities, CDI, in the Poynting flux dominated
jets, and magnetic Kelvin-Helmholtz instabilities, KHI, in the
kinetically dominated ones}; see, e.g.][]{Ha11} makes the flow
eventually subsonic preventing the formation of any subsequent
collimated flow beyond some axial distance.

  In the case of KHI, it is known that the growth
rates of the unstable modes are always larger for smaller Mach
number jets, and also that a shear layer surrounding the jet reduces 
the growth rates of long, disruptive wavelengths \citep[see, e.g.,
][]{PH04,PM05}. This is related to the number of reflections at 
the jet/ambient interface that the waves suffer within a given time
or distance, which increases with increasing (magnetosonic) Mach angle
(decreasing Mach number). Thus, taking into account that the growth of
the unstable modes occurs at this interface \citep[][]{PC85}, the
larger the number of interactions is, the faster the growth of the wave
amplitude. Linear analysis of the CDI also leads to the conclusion
that the growth rates of the modes decrease (or equivalently their
growth lengths increase) with increasing flow velocity
\citep[][]{AL00}, i.e., with increasing magnetosonic Mach numbers for
constant fast magnetosonic speeds. Numerical experiments have also
shown that the CDI growth rates are also reduced in the case of
magnetized flows with parallel magnetic fields or flows shrouded by
(magnetized) winds \citep[see, e.g., ][]{HR02,MH07,Ha11}. Unfortunately, no 
studies of jet stability for the case of sheared, magnetised,
relativistic jets have been performed so far, but the aforementioned
results from simulations of CDI development in the jet/wind scenario
point in the same direction than for of sheared, non-magnetized, {relativistic} jets
(Perucho et al. 2005).  Hence, in an attempt to reduce the growth of
pinch instabilities in our simulations to allow the
injected models to reach a steady state, we have
introduced shear layers of different widths depending on the
model. The properties of these shear layers are described in 
Sect~\ref{s:param} and Table~\ref{t:t1}. {Our results
  prove their stabilization effect of the CDI in Poynting-flux dominated
  jets.} Broadly speaking, the width of the shear layer has been set
as the minimum one allowing the injected model to establish an steady
solution along several (typically two or three) spatial
periods\footnote{In a recent paper, \cite{KB15} have
  focussed in the stability of (non-relativistic) magnetized jets that 
  carry no net electric current and do not have current sheets. The
  introduction of current-sheet-free magnetic fields significantly
  improves jet stability relative to unmagnetized jets or magnetized
  jets with current sheets at their surface. Moreover, the
  introduction of shear \citep{KB16} has also a strongly stabilizing
  effect on various modes of jet instability. Our results based on
  simulations of sheared jets without current sheets at their
  surfaces, extend (at least qualitatively) these results to the
  relativistic regime.}.

  The fact that we had to introduce such a shear layer to stabilize
our models against the growth of pinching modes is interesting in
itself. Taking into account that the transient phase between the
initial state and the desired steady one is not specially severe, the
fast growth of pinch instabilities reflects the difficulties of these
models (as mentioned before, especially those corresponding to cold,
Poynting flux dominated jets) to establish long-term, steady flows as
those inferred in AGN jets at parsec scales.

  {Besides this stabilizing effect, the introduction
of the shear layer has two additional effects: { 1) since the internal
structure of the jets is a consequence of the saturation of pinch
modes, adding a shear layer changes the internal structure of the
idealized top-hat jet models, and 2)} the introduction of
the shear layer also changes the original values of the parameters of
the injected models, listed in Table~\ref{t:t1} (see
Sect.~\ref{s:param}).}  Table~\ref{t:t2} contains average values of
several relevant quantities for the steady models of the jets analyzed
in this work as well as relative maximum variations of these
quantities along the jet.  Two rows per model are displayed. The first
row (with the corresponding models labeled ``s'') shows values in the
jet spine, defined as the region of the jet around the axis with jet
mass fraction $f \ge 0.995$. The second row (with models labeled
``j'') shows averages for the whole jet ($f>0$). Note that the average
values for the jet spine are not the same as those displayed in
Table~\ref{t:t1} for the corresponding model. This is because the
values shown in this table are defined at the injection point and
correspond to extreme values (i.e., maxima or minima) more than
average ones. In particular, the values of the rest-mass density, flow
Lorentz factor and magnetic pitch angle of the spine of steady models
are, respectively, $(3.9 \pm 0.4) \times 10^{-3}$, $3.4 \pm 0.2$, and
$52.3 \pm 0.5^\circ$, instead of the values at injection $5 \times
10^{-3}$, $3.2$ and $45^\circ$.  On the other hand, average values in
models j are contaminated by the presence of the shear layer. In these
models, the values of the rest-mass density, flow Lorentz factor and
magnetic pitch angle are respectively $0.09 \pm 0.03$, $2.7 \pm 0.3$
and $50.1 \pm 0.8^\circ$. The average magnetization of {the s and j models}
are always smaller than the corresponding reference values in
Table~\ref{t:t1} {and} those of the specific internal energy are
also smaller, exception made of models PK02s, PK03s, KP06s, KP10s (i.e,
the spine jets of the models with the highest magnetization). As a
result of these variations, models s and j are slightly shifted in the
${\cal M}_{ms,j}$ - $1/\varepsilon_j$ plane with respect to their
parent models but still belong to the same family of models (hot,
kinetically dominated, Poynting flux dominated).  

%
\begin{figure*}
\begin{minipage}{180mm}
\begin{center}
\includegraphics[width=18.cm,angle=0]{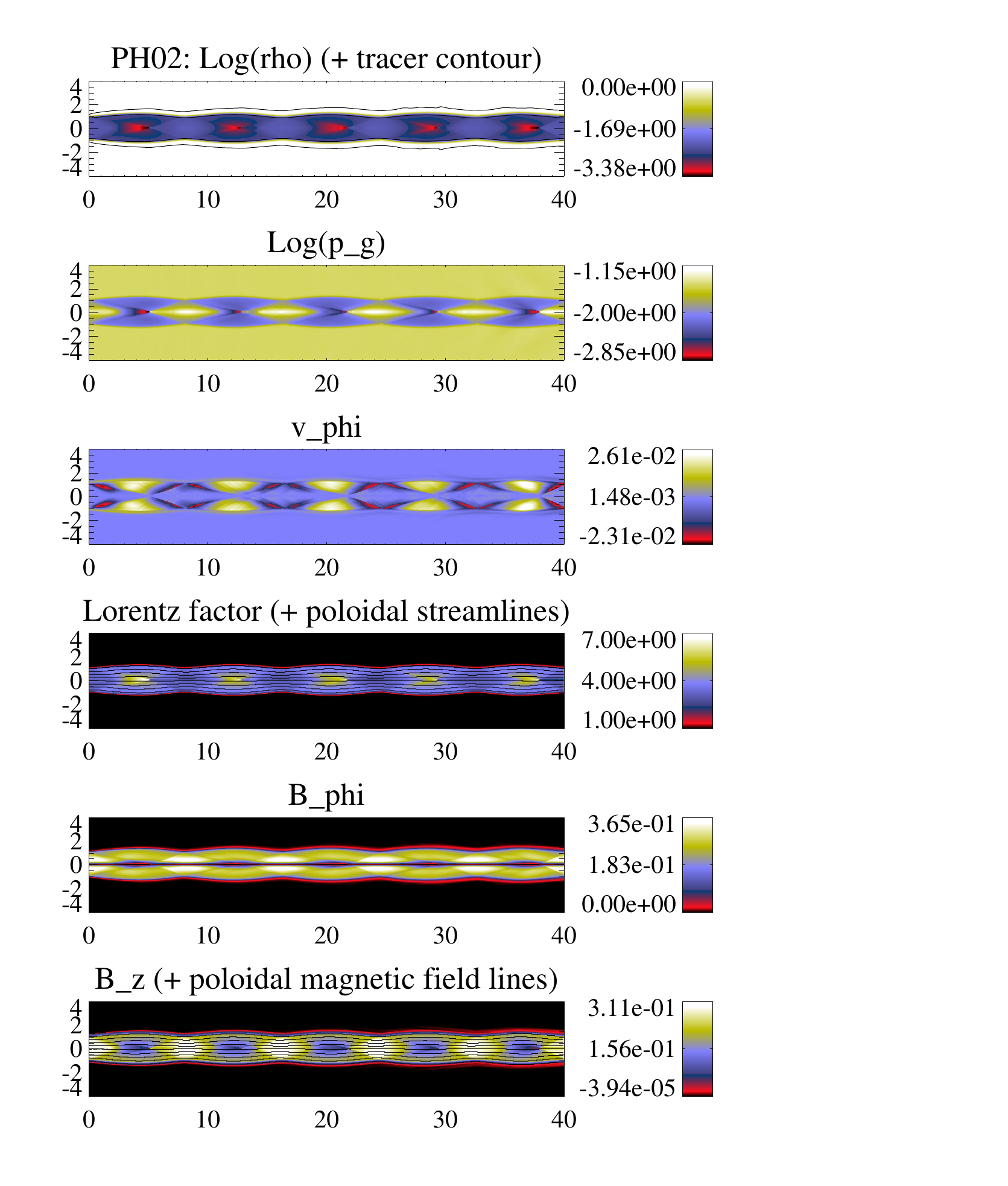}
\caption{Steady structure of the Poynting flux dominated, hot jet
  model PH02. From top to bottom, distributions of rest-mass density, gas
  pressure, toroidal flow velocity, flow Lorentz factor, and toroidal
  and axial magnetic field components once the steady
  state has been settled. Poloidal flow and magnetic field lines are
  overimposed onto the Lorentz factor and axial magnetic field panels,
  respectively. Two contour lines for jet mass fraction values 0.005 and 0.995
  are overplotted on the density panel.}
\label{f:PH02}
\end{center}
\end{minipage}
\end{figure*}
%

%
\begin{figure*}
\begin{minipage}{180mm}
\begin{center}
\includegraphics[width=18.cm,angle=0]{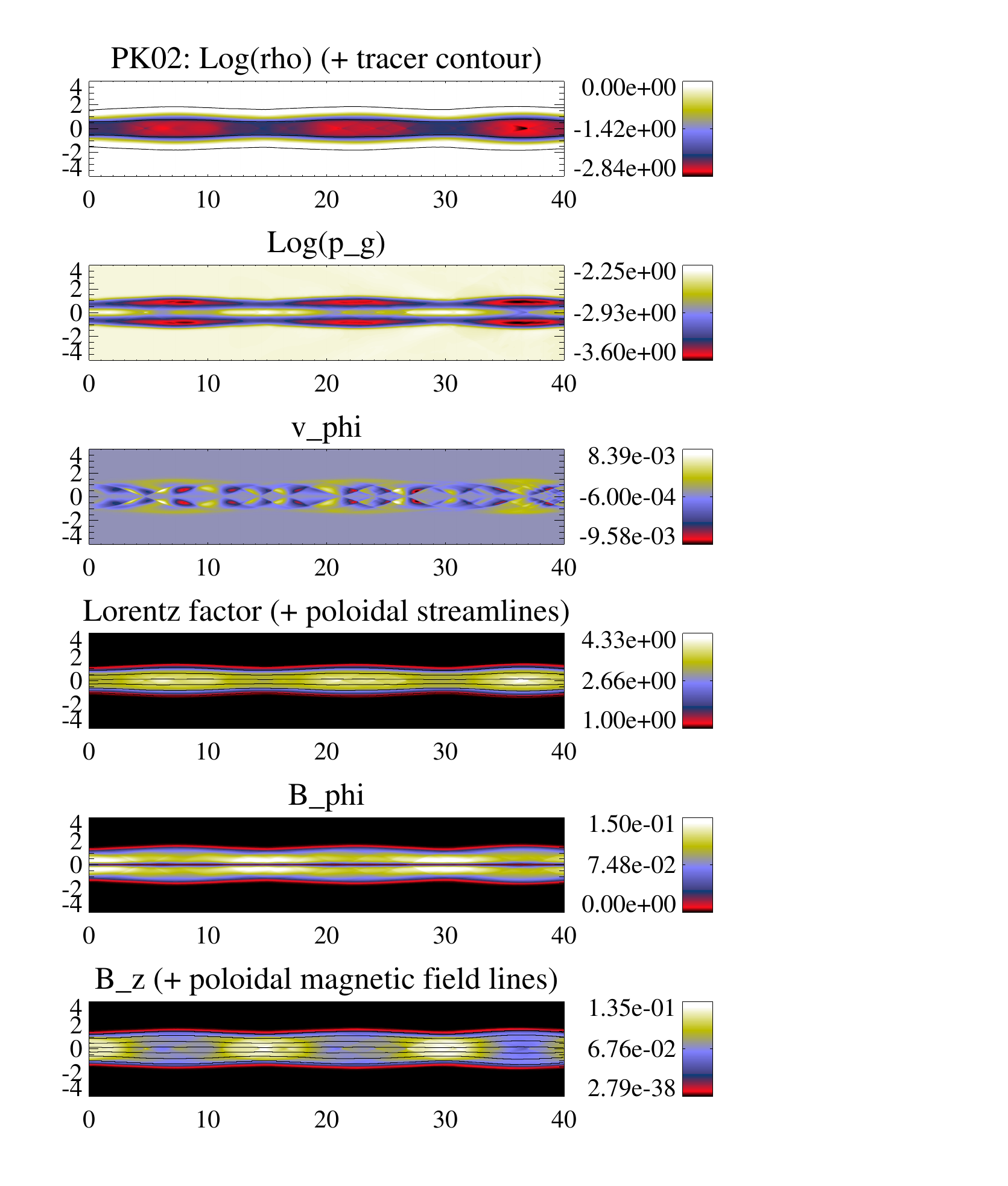}
\caption{Steady structure of the Poynting flux dominated jet model PK02. Panel
  distribution as in Fig.~\ref{f:PH02}.} 
\label{f:PK02}
\end{center}
\end{minipage}
\end{figure*}
%

%
\begin{figure*}
\begin{minipage}{180mm}
\begin{center}
\includegraphics[width=18.cm,angle=0]{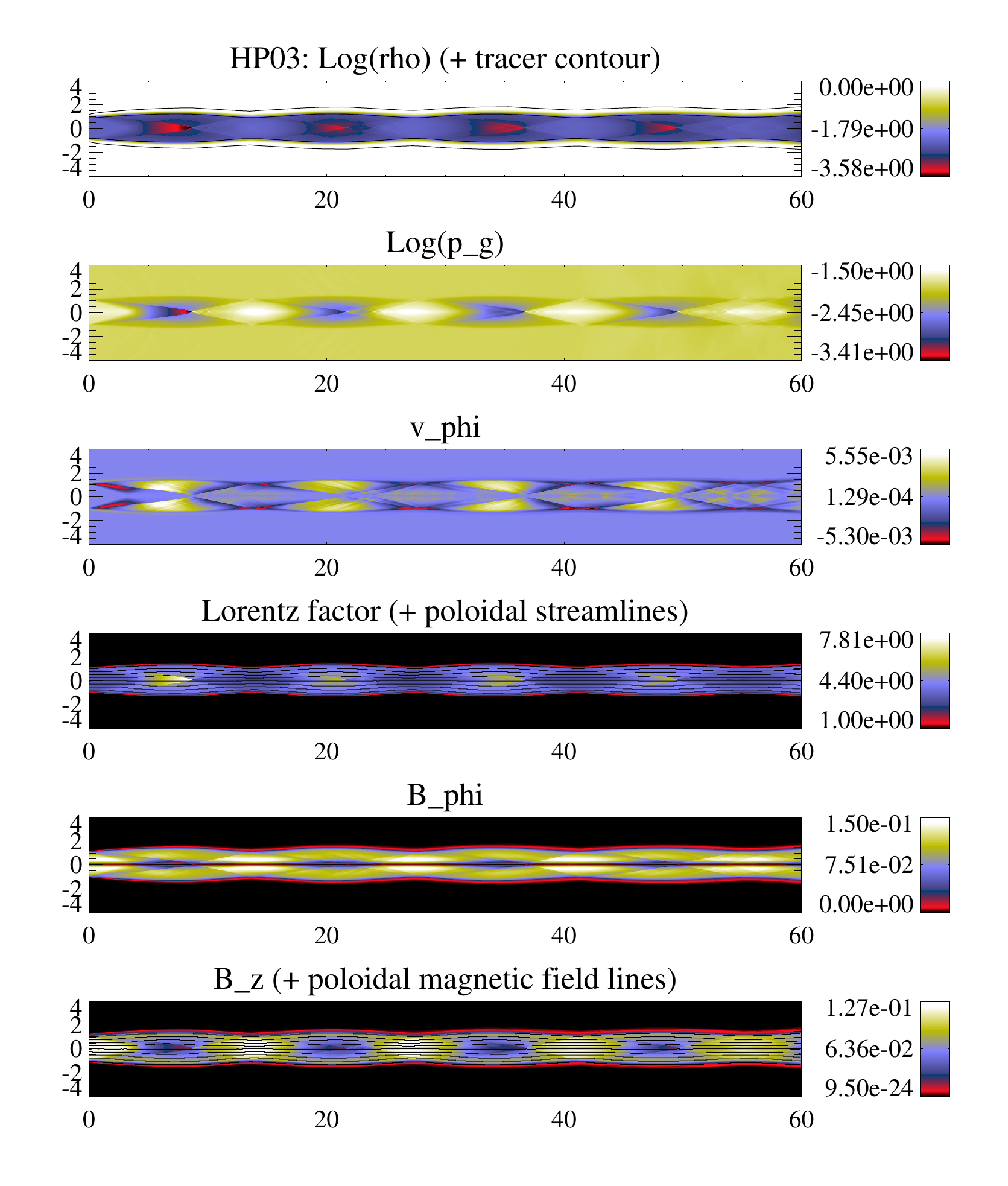}
\caption{Steady structure of the hot jet model HP03. Panel
  distribution as in Fig.~\ref{f:PH02}.} 
\label{f:HP03}
\end{center}
\end{minipage}
\end{figure*}
%

%
\begin{figure*}
\begin{minipage}{180mm}
\begin{center}
\includegraphics[width=18.cm,angle=0]{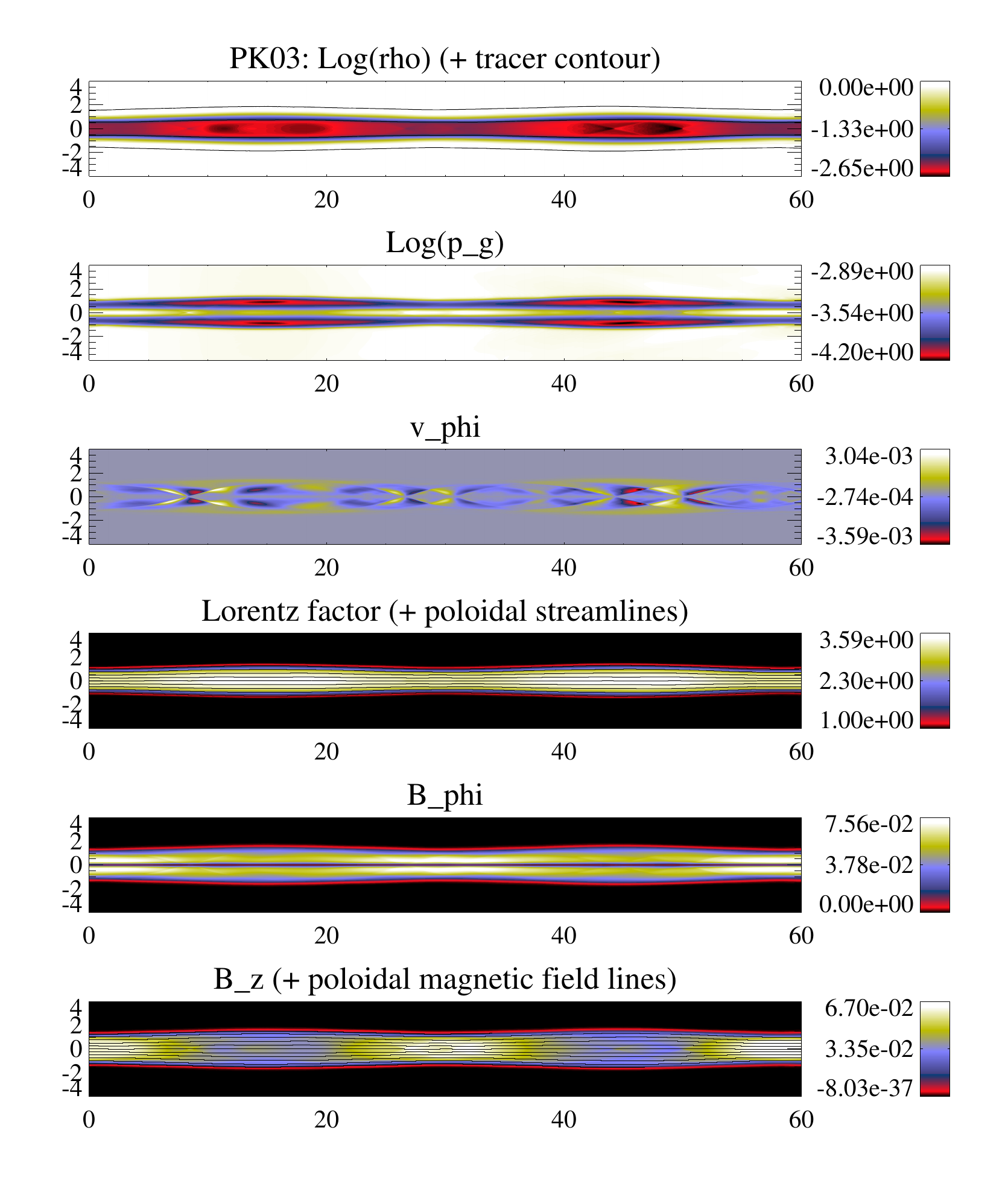}
\caption{Steady structure of the Poynting flux dominated jet model PK03. Panel
  distribution as in Fig.~\ref{f:PH02}.} 
\label{f:PK03}
\end{center}
\end{minipage}
\end{figure*}
%

%
\begin{figure*}
\begin{center}
\begin{minipage}{180mm}
\includegraphics[width=18.cm,angle=0]{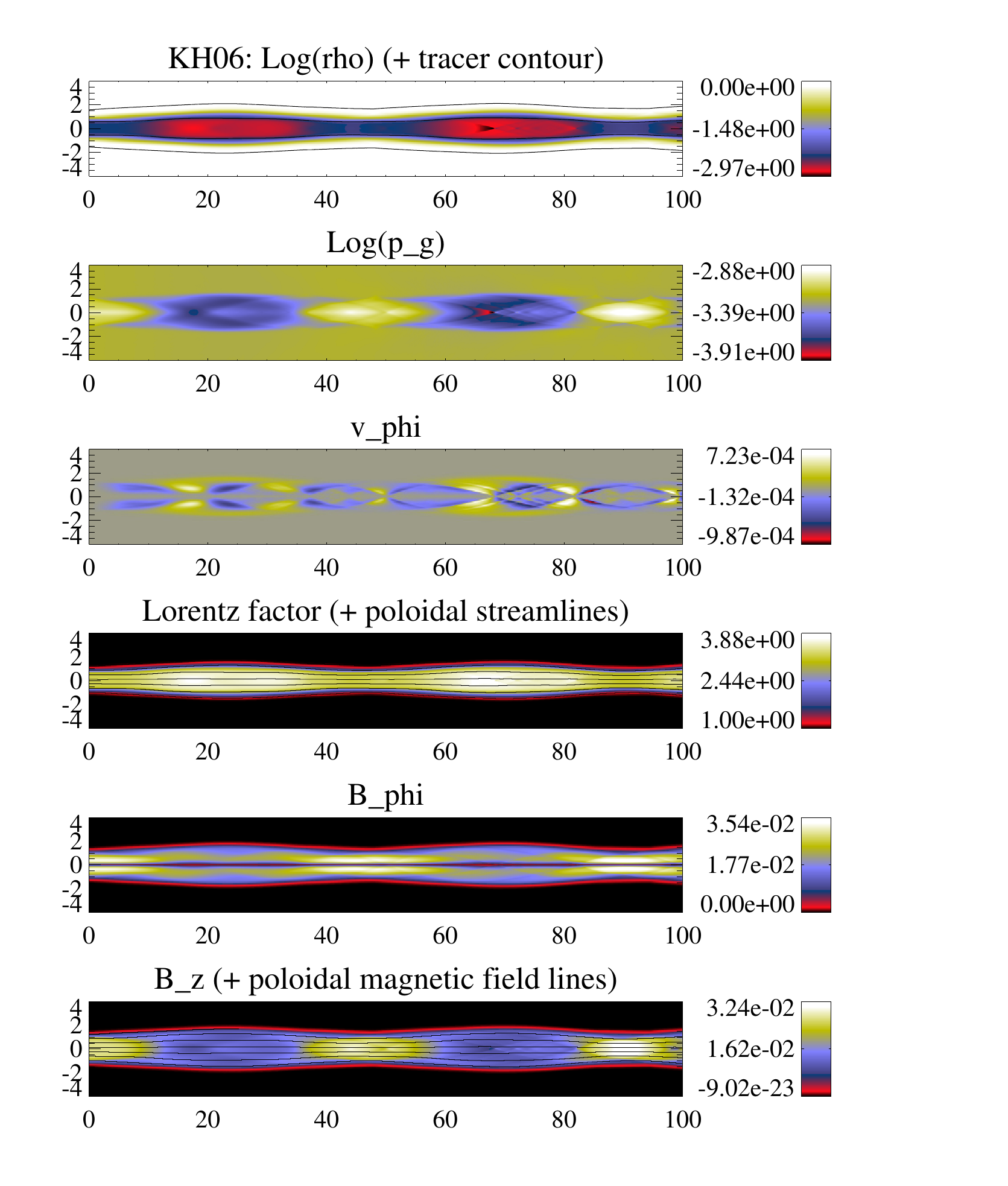}
\caption{Steady structure of the kinetically dominated jet model KH06. Panel
  distribution as in Fig.~\ref{f:PH02}. Note that the axial scale
has been compressed by a factor of 2 with respect to the radial one.} 
\label{f:KH06}
\end{minipage}
\end{center}
\end{figure*}
%

%
\begin{figure*}
\begin{center}
\begin{minipage}{180mm}
\includegraphics[width=18.cm,angle=0]{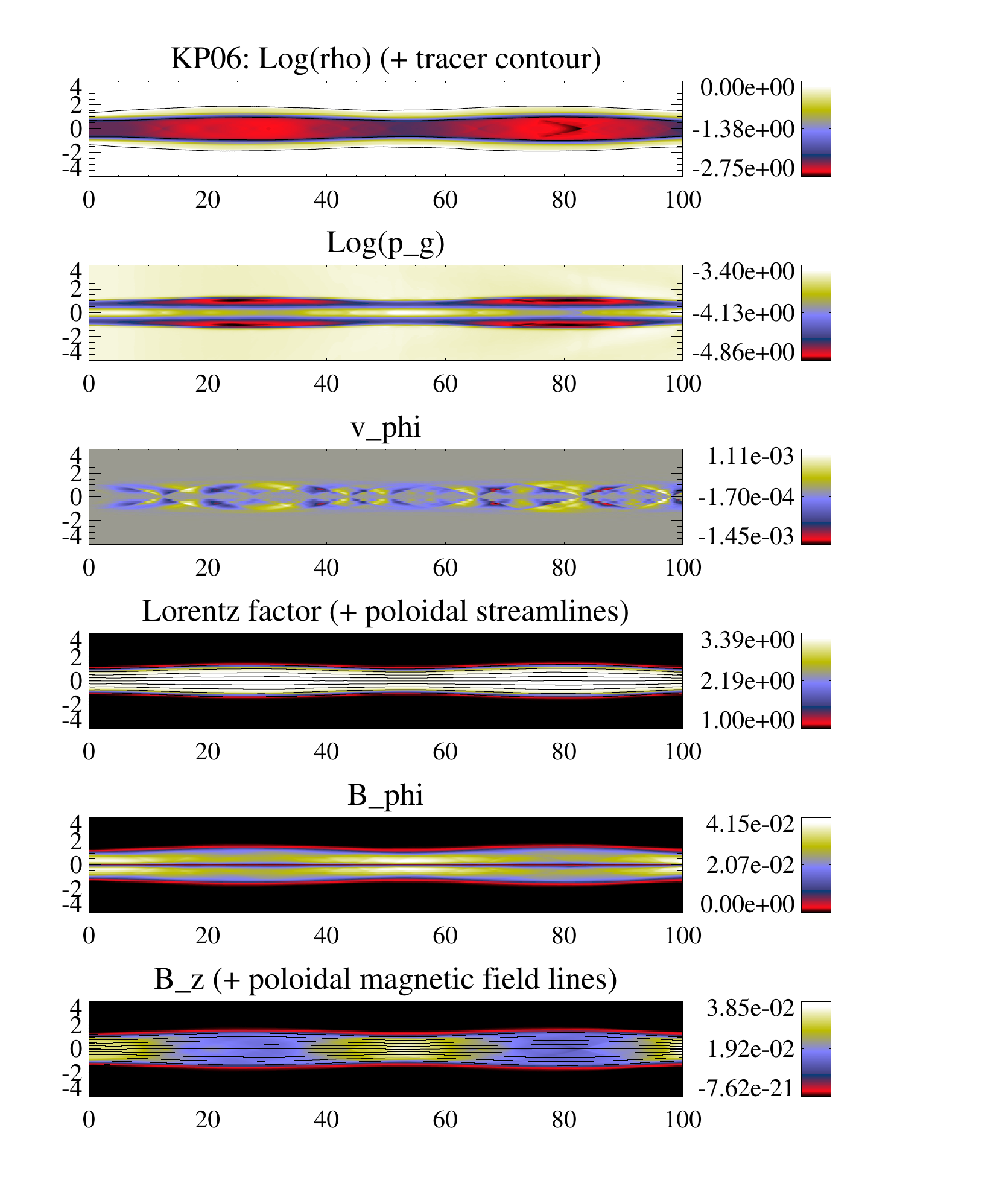}
\caption{Steady structure of the kinetically dominated, highly magnetized jet model KP06. Panel
  distribution as in Fig.~\ref{f:PH02}. Note that the axial scale
has been compressed by a factor of 2 with respect to the radial one.} 
\label{f:KP06}
\end{minipage}
\end{center}
\end{figure*}
%

%
\begin{figure*}
\begin{center}
\begin{minipage}{180mm}
\includegraphics[width=18.cm,angle=0]{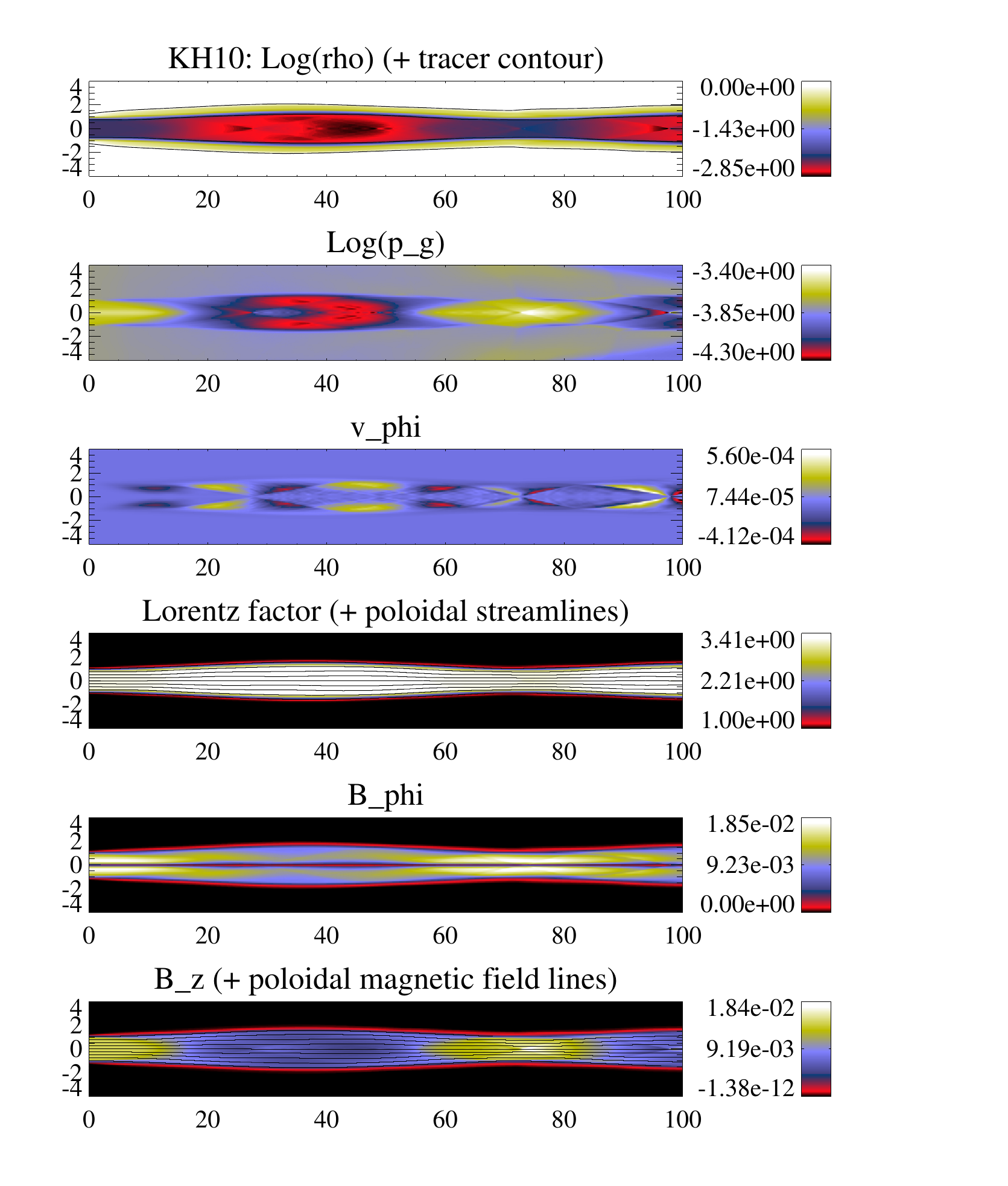}
\caption{Steady structure of the kinetically dominated jet model KH10. Panel
  distribution as in Fig.~\ref{f:PH02}. Note that the axial scale
has been compressed by a factor of 2 with respect to the radial one.} 
\label{f:KH10}
\end{minipage}
\end{center}
\end{figure*}
%

%
\begin{figure*}
\begin{center}
\begin{minipage}{180mm}
\includegraphics[width=18.cm,angle=0]{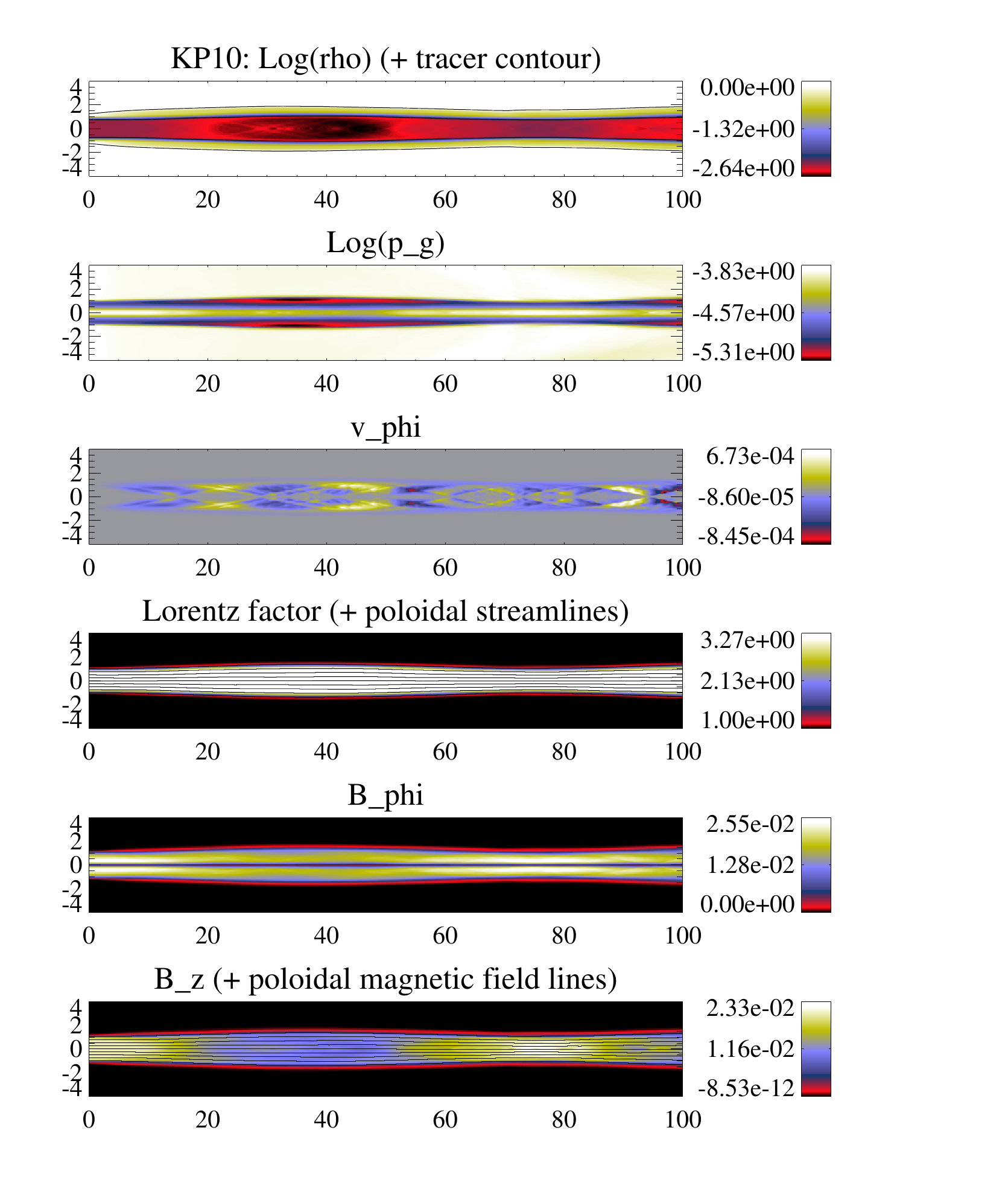}
\caption{Steady structure of the kinetically dominated, highly magnetized jet model KP10. Panel
  distribution as in Fig.~\ref{f:PH02}. Note that the axial scale
has been compressed by a factor of 2 with respect to the radial one.} 
\label{f:KP10}
\end{minipage}
\end{center}
\end{figure*}
%

%
\begin{table*}
 \centering
 \begin{minipage}{165mm}
\small
  \caption{Averaged values and relative variations along the jet of
    the quantities defining the steady models. $\Delta z$ stands for
    the wavelength associated to the jet oscillation along the
    axis. Remaining symbols, as defined in Sect.~\ref{s:param}.}
  \begin{tabular}{lccccccccccccc}
  \hline \\
  Model$^a$ & $\bar{R} \,[R_j]$ &  $\displaystyle{\frac{\Delta
      R}{\bar{R}}}$ & $\bar{\rho} \, [\rho_a]$ & $\displaystyle{\frac{\Delta \rho}{\bar{\rho}}}$ &
  $\bar{\varepsilon} \, [c^2]$ & $\displaystyle{\frac{\Delta
  \varepsilon}{\bar{\varepsilon}}}$ & 
  $\bar{\beta}$ & $\displaystyle{\frac{\Delta
  \beta}{\bar{\beta}}}$ & $\bar{\phi} \, [^\circ]$ & $\displaystyle{\frac{\Delta \phi}{\bar{\phi}}}$ & $\bar{W}$ &
  $\displaystyle{\frac{\Delta W}{\bar{W}}}$ & $\Delta z^b \,[R_j]$ \\
\\
  \hline
\\
  PH02s & $1.00$ & $0.26$ & $0.0034$ & $0.74$ & $9.00\phantom{00}$ & $0.24$ &
  $2.67$ & $0.02$ & $51.71$ & $0.19$ & $3.69$ & $0.22$ &
  $\phantom{0}8.0$ \\
  PH02j & $1.20$ & $0.25$ & $0.060\phantom{0}$ & $0.43$ & $6.56\phantom{00}$ & $0.14$ &
  $2.25$ & $0.06$ & $51.19$ & $0.18$ & $3.18$ & $0.24$ &
  $\phantom{0}8.0$ \\
  PK02s & $0.64$ & $0.23$ & $0.0045$ & $0.64$ & $0.76\phantom{00}$ & $0.24$ & $6.19$ & $0.21$ & $52.05$
  & $0.15$ & $3.34$ & $0.09$ & $14.5$ \\
  PK02j & $1.22$ & $0.20$ & $0.16\phantom{00}$ & $0.40$ &
          $0.23\phantom{00}$ & $0.11$ & $6.27$ & $0.12$ & $48.55$ & $0.14$
          & $2.29$ & $0.13$ & $14.5$ \\
  HP03s & $1.03$ & $0.29$ & $0.0032$ & $0.91$ & $8.31\phantom{00}$
  & $0.28$ & $0.44$ & $0.02$ & $52.25$ & $0.21$
  & $3.77$ & $0.27$ & $13.8$ \\
  HP03j & $1.25$ & $0.24$ & $0.050\phantom{0}$ & $0.52$ & $5.93\phantom{00}$ & $0.19$ & $0.37$ & $0.11$ &
         $51.35$ & $0.21$ & $3.24$ & $0.28$ & $13.8$ \\
  PK03s & $0.65$ & $0.23$ & $0.0046$ & $0.63$ & $0.20\phantom{00}$
  & $0.25$ & $6.08$ & $0.25$ & $52.05$ & $0.13$ 
  & $3.24$ & $0.03$ & $30.0$ \\
  PK03j & $1.23$ & $0.21$ & $0.16\phantom{00}$ & $0.44$ &
          $0.060\phantom{0}$ & $0.13$ & $6.13$ & $0.18$ & $48.61$ & $0.19$
          & $2.24$ & $0.09$ & $30.0$ \\
   KH06s & $1.09$ & $0.46$ & $0.0030$ & $1.20$ & $0.37\phantom{00}$ &
   $0.43$ & $0.40$ & $0.45$ & $
   55.43$ & $0.31$ & $3.45$ & $0.17$ & $38.0$ \\
   KH06j & $1.57$ & $0.38$ & $0.037\phantom{0}$ & $0.84$ &
   $0.201\phantom{0}$ & $0.25$ &
   $0.25$ & $0.28$ & $54.22$ & $0.28$ & $2.60$ & $0.23$ & $38.0$ \\
   KP06s & $0.82$ & $0.24$ & $0.0041$ & $0.76$ & $0.042\phantom{0}$
   & $0.24$ & $7.69$ & $0.39$ & $52.82$ & $0.20$
   & $3.20$ & $0.02$ & $53.0$ \\
   KP06j & $1.27$& $0.30$ & $0.097\phantom{0}$ & $0.41$ &
   $0.019\phantom{0}$ & $0.13$ &
   $6.73$ & $0.22$ & $50.00$ & $0.20$ & $2.42$ & $0.08$ & $53.0$ \\
   KH10s & $0.96$ & $0.47$ & $0.0036$ & $0.97$ & $0.11\phantom{00}$
   & $0.33$ & $0.41$ & $0.44$ & $53.90$ & $0.26$ &$3.25$ & $0.05$ & $74.0$ \\
   KH10j & $1.37$ & $0.40$ & $0.065\phantom{0}$ & $0.78$ &
          $0.058\phantom{0}$ & $0.17$ & $0.32$ & $0.31$ & $51.67$ & $0.25$
          & $2.54$ & $0.14$ & $70.0$ \\
   KP10s & $0.88$ & $0.32$ & $0.0042$ & $0.62$ & $0.016\phantom{0}$
   & $0.25$ & $8.96$ & $0.38$ & $51.74$ & $0.19$ &
   $3.19$ & $0.00$ & $76.0$ \\
   KP10j & $1.28$ & $0.27$ & $0.071\phantom{0}$ & $0.51$ &
          $0.0084$ & $0.14$ & $6.83$ & $0.20$ & $49.49$ & $0.20$ &
          $2.48$ & $0.08$ & $72.0$ \\
\\
\hline
\end{tabular}
\label{t:t2}
\\
$^a$ Two rows per model are displayed. The
first row (label s) shows values in the jet spine, defined as
the region of the jet around the axis with jet mass fraction $f \ge 0.995$. The
second row (label j) shows averages for the whole jet ($f>0$). \\
$^b$ $\Delta z$ stands for the wavelength of the jet oscillation along
the jet axis.
\end{minipage}
\end{table*}
%

  The internal structure of the jets will be now analyzed and compared 
with the help of Figs.~\ref{f:PH02} to \ref{f:KP10} and the results {given in
Table~\ref{t:t2}}. 

\begin{enumerate}

\item The models with a richer internal structure are those
  dominated by the internal energy, i.e., those in the hot models
  region or its neighbourhood (i.e., Poynting-flux dominated jets with 
  relatively small magnetization), PH02 and HP03. In these cases, the
  models have a substantial amount of internal energy which is
  efficiently converted into kinetic energy at jet expansions and back
  to internal energy at recollimation shocks. These models present the largest
  variations in flow Lorentz factor. The maximum Lorentz factor in
  model PH02 is 7.0 (2.19 times its initial value; see Fig.~\ref{f:PH02})
  and the change in the average values is of 22-24\% (see
  Table~\ref{t:t2}). In the case of model HP03, the maximum Lorentz
  factor is 7.81 (2.44 times its initial value) 
  and the change in the average values is of 27-28\%. { Associated
    to these variations in the flow Lorentz factor is the variation of
    the internal energy density along the axis. In the case of model PH02,
  this variation is a factor of $\approx 50$. For model HP03, it is a
  factor of $\approx 80$.}

\item Kinetically dominated jets (KH06 to KP10) are dominated
  by the rest-mass energy density and the inertia of the flow in these
  models is very large. In these models, specially in the colder ones
  (KH10, KP10), there is not much internal nor magnetic energy to be
  converted into kinetic one and the jets have no internal
  structure. In models KH10 and KP10 the maximum Lorentz factor is only
  1.02-1.06 times the initial value (see Figs.~\ref{f:KP06} and
  \ref{f:KP10}) and the change in the average values on the spine of
  the jets is smaller than 2\% (8\% including the shear layer; see
  Table~\ref{t:t2}). 

\item Models PK02, PK03 are Poynting-flux dominated models with 
  high magnetization ($\beta_j = 10$) and large magnetic pitch angles
  ($\phi_j =45^\circ$). In these cases, the width of the shear layer
  required to reduce the growth rate of the magnetic pinch modes could
  affect the internal structure of the models (see the discussion at
  the beginning of this section).

\item {The wavelength of the jet oscillation, $\Delta z$ (see
  Table~\ref{t:t2}), increases with increasing magnetosonic Mach
  number, as expected from the decrease of the Mach angle. This
  happens both for constant specific internal energy (and decreasing
  magnetization), and constant magnetization (and decreasing specific
  internal energy). Finally, for constant magnetosonic Mach number,
  this wavelength increases for decreasing specific internal energy
  (or increasing magnetization)}. Kinetically dominated jets 
  tend to have the longest wavelengths. In the models with the highest
  specific internal energies the oscillation of the jet leads to a
  series of { recollimation} shocks of the same periodicity. The 
  angle formed by these { conical} shocks with the jet axis is of $\approx
  18^\circ$ for model PH02 and $\approx 12^\circ$ for model
  HP03 (see below). 

\item Finally, the change in magnetic pitch angle for all models is limited to a mere
  25-26\% corresponding to a variation of $\pm 6^\circ$ around the
  average value.

\end{enumerate}

\subsection{Standing shocks}
\label{ss:is}

  Standing oblique discontinuities { (recollimation shocks)} can be
  distinguished in some of the jets, specially in hot models (PH02,
  HP03) and, to a lesser extent, in colder, low magnetization models
  (KH06, KH10). 

  The characterization of discontinuities in a magnetized fluid can be
done through the jumps of the different variables across them
\citep[see the books by][for a complete discussion]{Li67,An89}. Taking
into account that across a shock the specific entropy increases,
$s_b>s_a$ (where subscripts $a$ and $b$ refer to the state ahead and 
behind the shock, respectively), the compressibility assumptions
$\partial (h/\rho)/\partial p |_s <0$, $\partial (h/\rho)/\partial s
|_p >0$, $\partial^2 (h/\rho)/\partial p^2 |_s >0$, verified by the
ideal gas describing the jet fluid\footnote{These
  compressibility assumptions, as {named by} \cite{Li67}, qualify the
  equation of state describing the matter as convex \citep{IC13}.},
lead to $p_b>p_a$, $\rho_b>\rho_a$, $h_b>h_a$ ($(h/\rho)_b <
(h/\rho)_a$). Additionally, the magnetic pressure increases across a
fast magnetosonic shock, $p_{m,b}>p_{m,a}$, and decreases across a slow 
magnetosonic shock, $p_{m,b}<p_{m,a}$. Finally,
the increase in density across a shock, coincides with a decrease in
the flow speed (or in the spatial component of the four-velocity) in
the shock rest-frame. 

\begin{figure}
\includegraphics[width=9.cm,angle=0]{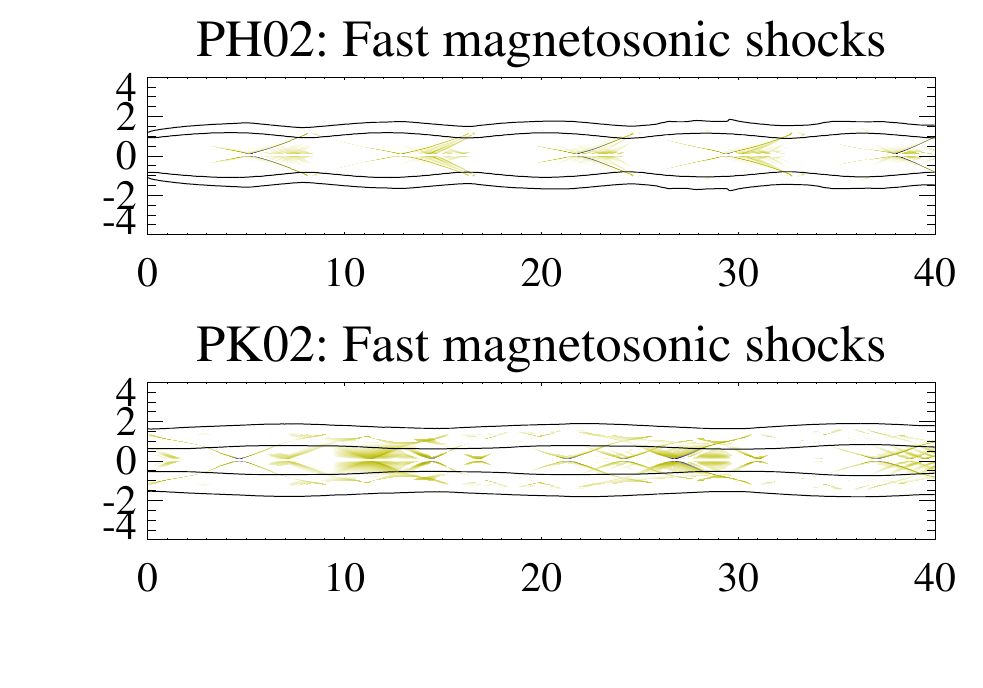}
\caption{Fast magnetosonic shocks in models PH02 and PK02 (${\cal
    M}_{ms, j}= 2.0$).} 
\label{f:PH02-PK02_fmshocks}
\end{figure}

\begin{figure*}
\includegraphics[width=13.cm,angle=0]{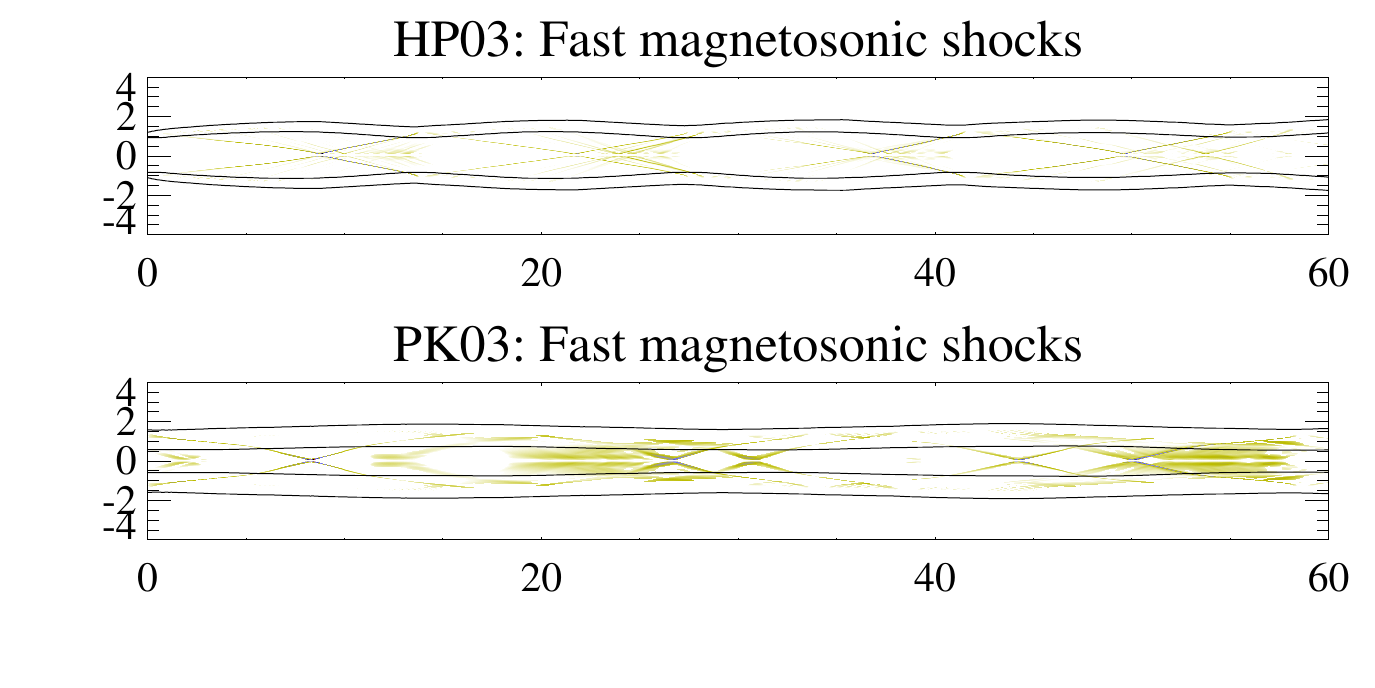}
\caption{Fast magnetosonic shocks in models HP03 and PK03 (${\cal
    M}_{ms, j}= 3.5$).} 
\label{f:HP03-PK03_fmshocks}
\end{figure*}

\begin{figure*}
\includegraphics[width=13.cm,angle=0]{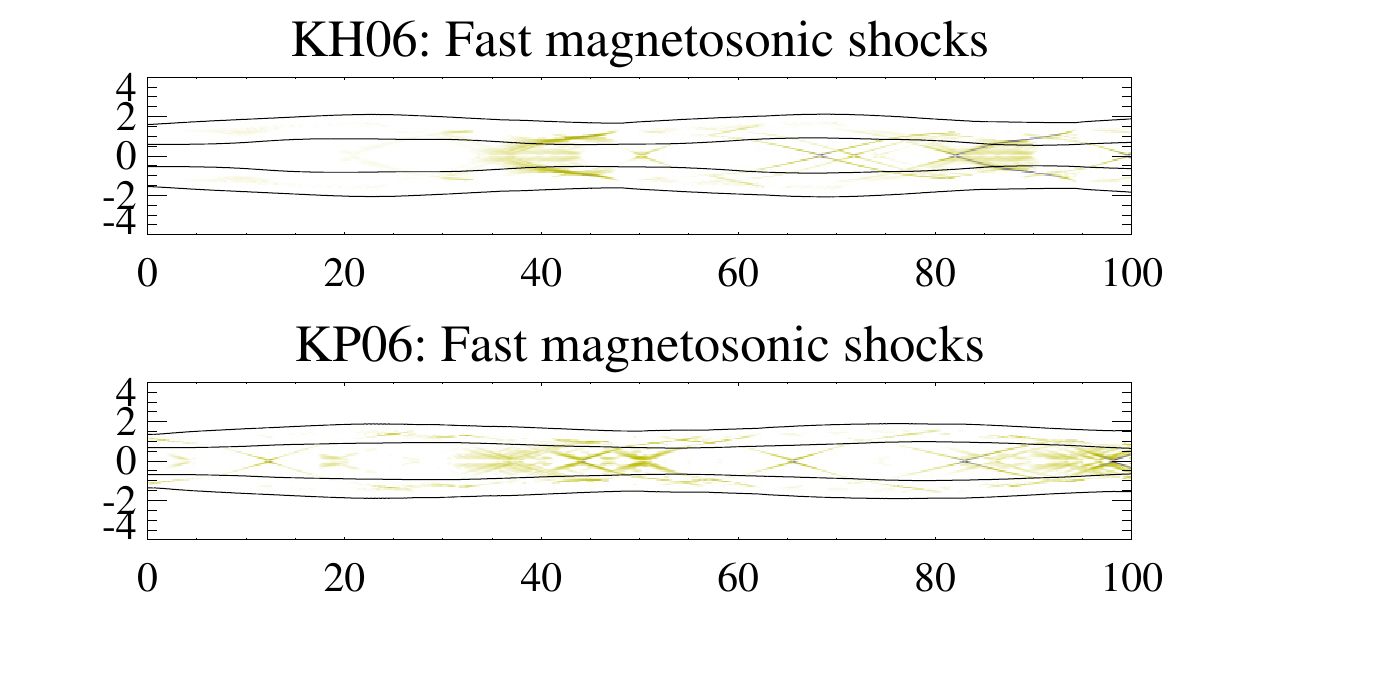}
\caption{Fast magnetosonic shocks in models KH06 and KP06 (${\cal
    M}_{ms, j}= 6.0$). Note that the axial scale
has been compressed by a factor of 2 with respect to the radial one.} 
\label{f:KH06-KP06_fmshocks}
\end{figure*}

\begin{figure*}
\includegraphics[width=13.cm,angle=0]{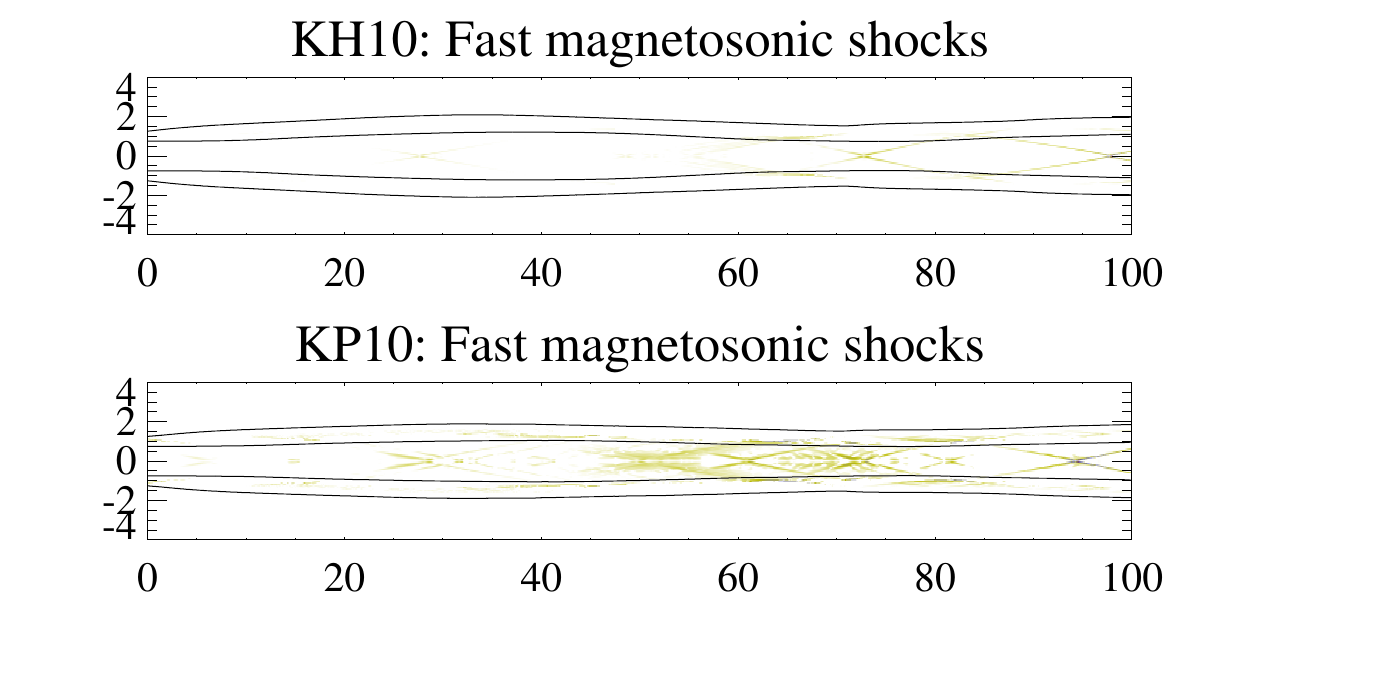}
\caption{Fast magnetosonic shocks in models KH10 and KP10 (${\cal
    M}_{ms, j}= 10.0$). Note that the axial scale
has been compressed by a factor of 2 with respect to the radial one.} 
\label{f:KH10-KP10_fmshocks}
\end{figure*}

  In a unsteady flow, identifying the states ahead and behind a shock
can be a difficult task. However in our steady jet models, fluid
particles enter the hypothetical discontinuities from the left
(smaller axial coordinate $z$) and leave them from the right (larger
$z$). With all this in mind we have built a detector of superfast
magnetosonic shocks based on the gradients of the thermal and magnetic 
pressures and the divergence of velocity. The results are shown in
Figs.~\ref{f:PH02-PK02_fmshocks}-\ref{f:KH10-KP10_fmshocks}. These
figures show the map of $d = |\Delta v_-| \Delta p_+ \Delta p_{m,+}$,
where $\Delta v_-$ is defined as the divergence of the
three-velocity if it is negative, and zero otherwise, and $\Delta p_+$
and $\Delta p_{m,+}$, as the corresponding gradients of the
thermal, $p$, and magnetic, $p_m$, pressure if the $z$ components of
the gradient are positive (and zero otherwise). According to our
analysis, the internal
structures seen in the color panels of the different models
(Figs.~\ref{f:PH02}-\ref{f:KP10}) are identified as fast magnetosonic
shocks in Figs.~\ref{f:PH02-PK02_fmshocks}-\ref{f:KH10-KP10_fmshocks}.

Some conclusions can be drawn from these figures. Models PH02 and HP03
(corresponding to the hottest jets studied and with the thinnest shear
layer) and to a lesser extent models KH06 and KH10 (kinetically
dominated jet models) { display a series of periodic recollimation
  shocks} associated with 
the jet sideways oscillations. In the remaining models (Poynting-flux
dominated models with the widest shear layer), the primary shocks
associated with the oscillations of the jet are weaker and dilute in a
more complex structure of shocks (and compression waves). All the
standing shocks in a jet form the same angle with respect to the jet
axis and there is a clean correlation between the shock angle (the
angle formed by the conical shock and the jet axis) and the
magnetosonic Mach number (see Table~\ref{t:shockangle}). 

{Finally, let us note that none of our models display
  Mach disks. This kind of structures replace conical shocks when the
  shocks are strong enough. Since in our simulations the strength of the
  standing shocks depends on the jet overpressure factor, we would
  expect to find Mach disks for larger overpressure factors.}

%
%

%
%

%
%

%
%

%
\begin{table}
\centering
\caption{Shock angle, $\phi_s$, as a function of the
    magnetosonic Mach number, ${\cal M}_{ms}$. The Mach angle, $\phi_M
    = \arctan(1/{\cal M}_{ms})$ is also shown for comparison.}
\begin{tabular}{rrr}
\hline \\
${\cal M}_{ms}$ & $\phi_M$\,[$^\circ$] & $\phi_s$\,[$^\circ$]\\
\\
  \hline
\\
 2.0  & 26.6 & 18 $\pm$ 1 \\
 3.5  & 15.9 & 13 $\pm$ 1\\
 6.0  & 9.5& 8 $\pm$ 1\\
 10.0 & 5.7& 6 $\pm$ 1\\
\\
\hline
\end{tabular}
\label{t:shockangle}
\end{table}
%

\subsection{Astrophysical applications}
\label{ss:apapp}

{The
correlation found in our simulations between the Mach angles, the
angles of the { recollimation} shocks and the separation between them (in
models with internal structure) allows us to estimate the Mach numbers
of parsec-scale jets with stationary components.}

  {Based on multifrequency VLBI observations in the period
May 2005-April 2007, \cite{FR13a} report on the properties of three
quasi-stationary components (B1, B2 and B3) located between 4 and 7
mas from the core of CTA~102. Taking these
components as standing shocks, the shock angle can be estimated as the
angle subtended by the jet diameter at a distance equal to the shock
separation. Our study is based on the 15GHz
  observations made on 
8th of June 2006 \citep[see Tables~A.3 and A.6, and Figs.~A.7
and A.8 in][for the component analysis at 15 GHz]{FR13a}.
Assuming the mean of the size of a pair of
consecutive components (estimated as the FWHM of the corresponding
fitting Gaussians) as a lower bound of the jet diameter between 
these components, and the corresponding shock separation
\citep[correcting projection effects for a viewing angle of 
2.6$^\circ$, ][]{JM05}, we find an upper bound for the Mach number of 31.8
for the flow between B3 and B2 components, and correspondingly of 34.2
for the flow between B2 and B1. For an estimated flow Lorentz 
factor of 10 \citep[from the apparent speeds estimated in the neighbour
regions A and D; see Table~3 and Fig.~13 of][]{FR13a}, the values
obtained for the Mach number upper bounds would fall into the
kinetically dominated jet region of the corresponding magnetosonic
Mach number-specific internal energy plot, indicating that the jet can
be kinetically dominated at distances $\approx 720-1260$ pc ($0.1$ mas
in projection $\approx$ $18$ pc deprojected) from
the central source (see below).} 

  {We can apply the same analysis to the quasi-steady
components inside the innermost jet regions in BL~Lacertae, recently
reported by \cite{GL16}. From the component separation and
component sizes (estimated again as the FWHM of the
corresponding Gaussian fits) at 43 GHz for the core and  
components Q1 and Q2 \citep[see Table~2 and Fig.~3 of][]{GL16}, and a
viewing angle of 8$^\circ$ \citep{JM05}, we can derive upper bounds for
the Mach numbers of the flow between the core and component Q1,
17.5, and between components Q1 and Q2, 19.4. For an estimated
flow Lorentz factor of 7 \citep{JM05}, the Mach numbers bounds lie
again in the kinetically dominated region but closer to the
hot/Poynting flux dominated region boundary, at $\approx 2.4$ pc
($0.1$ mas in projection $\approx$ $0.97$ pc
{deprojected}) from the central source. The parameters
used in \cite{GL16} to simulate the jet in BL~Lac corresponding to a hot,
low-magnetization jet, are consistent with our estimations.}

%
\begin{figure}
\includegraphics[width=8.6cm]{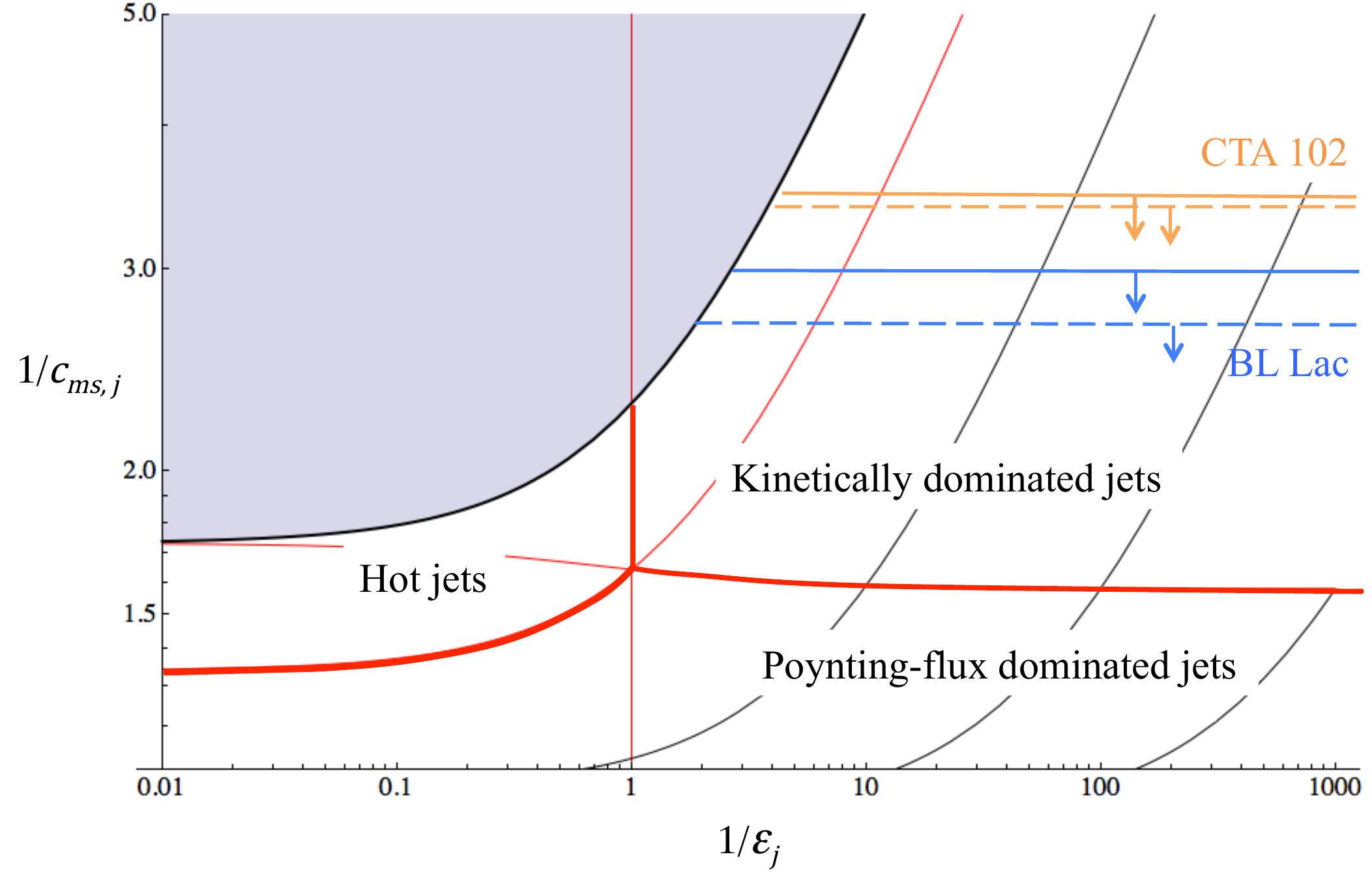} 
\caption{Classical magnetosonic Mach number estimates for the jets of
  CTA~102 and BL~Lacertae on the $1/c_{ms,j} - 1/\varepsilon_j$ plane
  and distribution of jet models according to the dominant energy
  type. The black and red lines appearing in this diagram have the
  same meaning as the corresponding ones in Fig.~\ref{f:f1}. {The
    yellow (blue) solid and dashed lines give maximum 
    values for the two estimates of the Mach number in the jet of
    CTA~102 (BL~Lac), as discussed in the text.}} 
\label{f:machdiag}
\vspace{1cm}
\end{figure}
%

  {The Mach number estimates for
CTA~102 and BL~Lacertae are shown in Fig.~\ref{f:machdiag},
equivalent to the top panel in Fig.~\ref{f:f1}, but replacing the
relativistic magnetosonic Mach number in the ordinate axis by the
inverse of the magnetosonic speed ($c_{ms}$; see Appendix~A), a very
good approximation of the classical 
counterpart of the magnetosonic Mach number for high Lorentz factor
flows. It is interesting to note that our estimates of the dominant
type of energy for the jets of BL~Lacertae and CTA~102 fits well
within the current paradigm of jet acceleration, in which
jets would form at some point in the hot/Poynting-flux dominated
region and would evolve towards the region of kinematically dominated
models. This trend would have to be confirmed with a larger sample of
sources with stationary components and estimates of the bulk flow Lorentz
factor and jet viewing angle. However we should note that the present
approach for the estimation of the dominant type of energy in the jet
is applicable only to jets displaying stationary components.}

\section{Summary and conclusions}

  The internal structure of eight superfast magnetosonic,
overpressured jet models has been analyzed. The injection parameters
of these models have been chosen to cover a wide region in the
magnetosonic Mach number - specific internal energy plane. The merit
of this plane is that models dominated by different kind of energies
(internal energy: hot jets; rest-mass energy: kinetically dominated
jets; magnetic energy: Poynting-flux dominated jets) occupy well
separated regions. The analyzed models also cover a wide range of
magnetizations. The rest of injection parameters (the
rest-mass density, the jet overpressure factor, the flow Lorentz
factor, and the flow azimuthal velocity -equal to zero-) are kept
constant. 

  Jets are injected in internal transversal equilibrium to minimize
the sideways perturbations once immersed in the ambient medium and to
obtain an internal structure as clean as possible. The transition
between the jet and the ambient medium is smoothed by means of a shear
layer of different widths to stabilize the models against the growth
of magnetic pinch instabilities. The conclusions of our analysis are listed
below. 

\begin{enumerate}

\item The models with a richer internal structure are those
  dominated by the internal energy, i.e., those in the {hot jets} 
  region or its neighbourhood (i.e., Poynting-flux dominated jets with  
  magnetizations larger than but close to 1). In these cases, the
  models have a substantial amount of internal energy which is
  efficiently converted into kinetic energy at jet expansions and back
  to internal energy at { recollimation} shocks. These models present the
  largest variations in flow Lorentz factor { and internal energy
  density along the axis}.

\item Conversely, in the {kinetically dominated jet} models there is
  not much internal nor 
  magnetic energy to be converted into kinetic one, {the jets} have
  no internal structure and the flow Lorentz factor is
  constant. Despite the large difference in magnetization, kinetically
  dominated models with the same magnetosonic Mach number have very 
  similar overall structure (jet oscillation, amplitude of variations,
  local jet opening angles,...). 

\item As a consequence of the magnetic pinch exerted by the
  toroidal magnetic field, {models with large magnetizations}
  concentrate most of their internal energy in a thin hot spine around
  the axis. The width of this spine is
  related with the location of the maximum toroidal field across the
  jet. 

\item {Poynting-flux dominated models with high magnetization} are
  prone to be unstable against magnetic pinch modes.

\item All the models present a jet oscillation with a characteristic
  wavelength that follows definite trends with specific internal
  energy, magnetosonic Mach number and magnetization. 

\item The { change in (average) magnetic pitch angle} is limited to
  few degrees around the average value. However, large local radial
  variations in the pitch angle can be expected from almost $0^\circ$
  close to the axis to values larger than the average at some
  intermediate radius. 

\item Despite the fact that the studied models are injected with pure
  axial flow velocities, all develop {small azimuthal velocities} (of the
  order of 2\% of the speed of light or smaller) as a result of the
  Lorentz force in axisymmetric converging/diverging flows. These
  speeds tend to be larger in those models where the jet oscillation
  has a larger amplitude.

\end{enumerate}

Despite its limitations,
the present study is the first attempt to identify the structural
ingredients (including the properties of { recollimation} shocks) that
characterize hot, Poynting-flux dominated and kinetically dominated,
relativistic jets. Our study is  of special relevance in the
interpretation of parsec-scale AGN jets. On one hand, our simulations confirm the
correlation between the Mach angles, the angles of the conical shocks
and the separation between them (in models with internal
structure) and allow us to estimate the magnetosonic Mach numbers of parsec-scale
jets with stationary components. {It should be noted,
  however, that our simulations are two-dimensional and that imperfect
  azimuthal symmetry of the ambient medium would disrupt the coherence
  of the standing shock pattern after few jet oscillations.} On the other hand, our study 
reveals that the presence of a significant toroidal component of the
magnetic field in these objects produces a complex transversal
structure with a central spine (extending up to the radius of the
maximum of the toroidal field) where the thermal pressure
(and hence the plasma internal energy) is close to its
maximum. {A layer with milder
(magnetic, thermal) pressure profiles that extends up to the outer
jet/ambient-medium transition layer wraps the central spine}. This complex profile in the
thermal energy distribution and the magnetic pitch angle {must} leave
their imprints in the total and polarized emission, which will be the
subject of a forthcoming paper. In that work we shall {analyze in
detail the emission properties of these models,} paying special attention
to the relative intensity of the components associated to the shocks
as a function of the viewing angle, to the tranversal structure  of
the jet and, in general, to the signatures of the magnetic field
structure in the polarized emission.

{Our results prove the stabilization effect of shear layers for the CDI
  in Poynting-flux dominated jets.} More interestingly, the stability
of Poynting flux dominated jets against pinch oscillations (and, in
particular, the role of the shear layer in the stabilization of these
flows) merits further exploration as a way to constrain the
magnetization parameter { and/or the magnetic field configuration}
in parsec-scale AGN jets.  

  Additional parameters should be explored, specially the overpressure
factor ({related with the formation of Mach disks})
and the magnetic field pitch angle, as well as new strategies to 
generate the steady models. In a recent paper, \cite{KP15}
describe a simple numerical approach to study the structure of steady
axisymmetric superfast-magnetosonic jets by means of one-dimensional
time dependent simulations by using $z$ (the axial cylindrical
coordinate) as the time coordinate. Although subject to a number of
approximations, the approach works well and could be used to generate
approximate steady solutions in a wider space of parameters.  

\acknowledgments
We thank Jos\'e M$^{\rm \underline{a}}$ Ib\'a\~nez and Alan P. Marscher for valuable comments that helped us to
improve the manuscript. J-MM and MP acknowledge financial support
from the Spanish Ministerio de Econom\'{\i}a y Competitividad (grants
AYA2013-40979-P, and AYA2013-48226-C3-2-P) and from the local
{Valencian Government} (Generalitat Valenciana, grant
Prometeo-II/2014/069). JLG acknowledges support from the Spanish
Ministerio de Econom\'{\i}a y Competitividad grant
AYA2013-40825-P. Calculations were carried out using the Altix UV1000
supercomputer {\it LluisVives} at the University of Valencia. 

\appendix

\section{Appendix A. Characteristic wavespeed diagrams for the RMHD}

\begin{figure*}
\begin{center}
\begin{minipage}{176mm}
\includegraphics[width=8.5cm,angle=0]{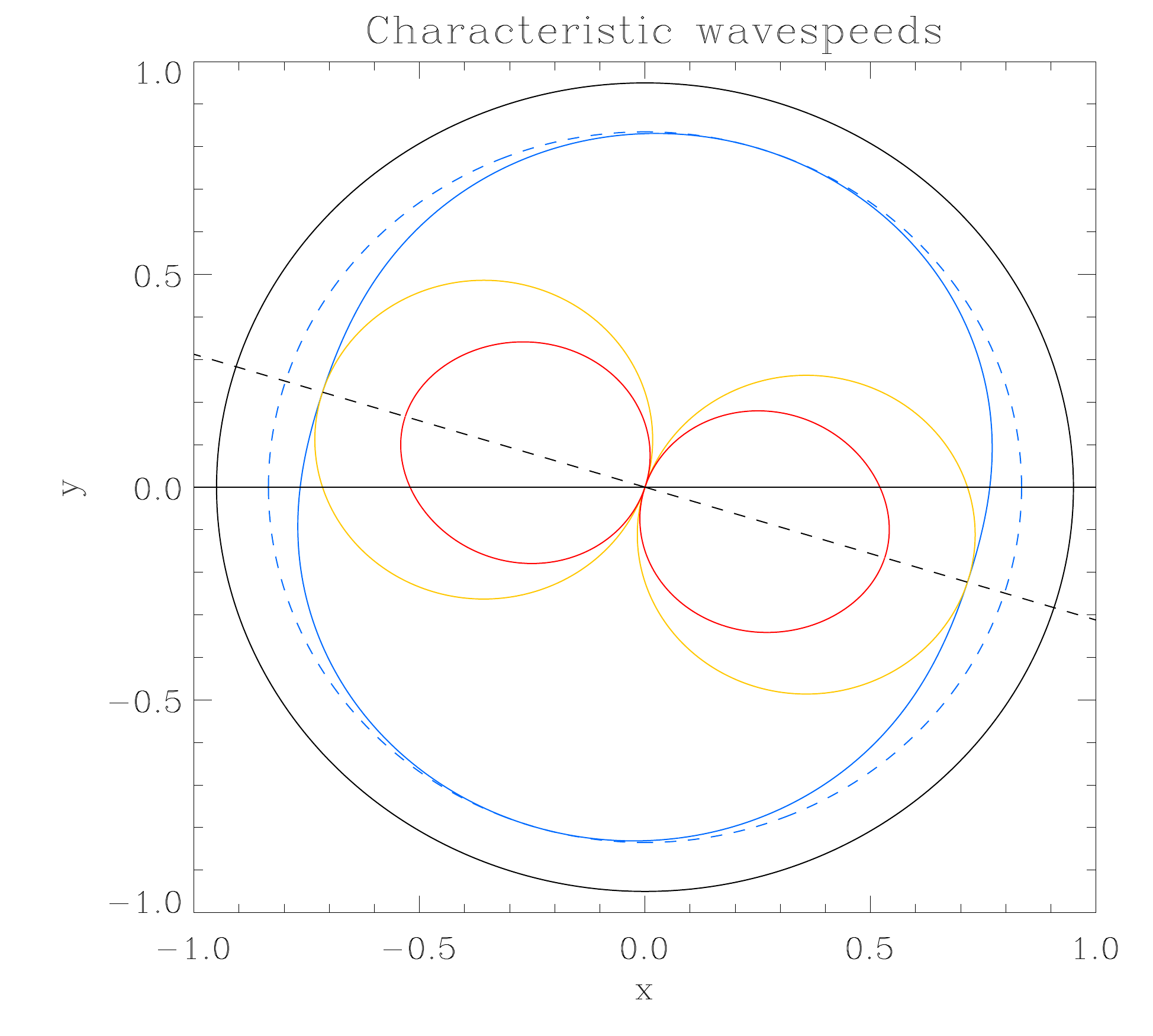} 
\includegraphics[width=8.5cm,angle=0]{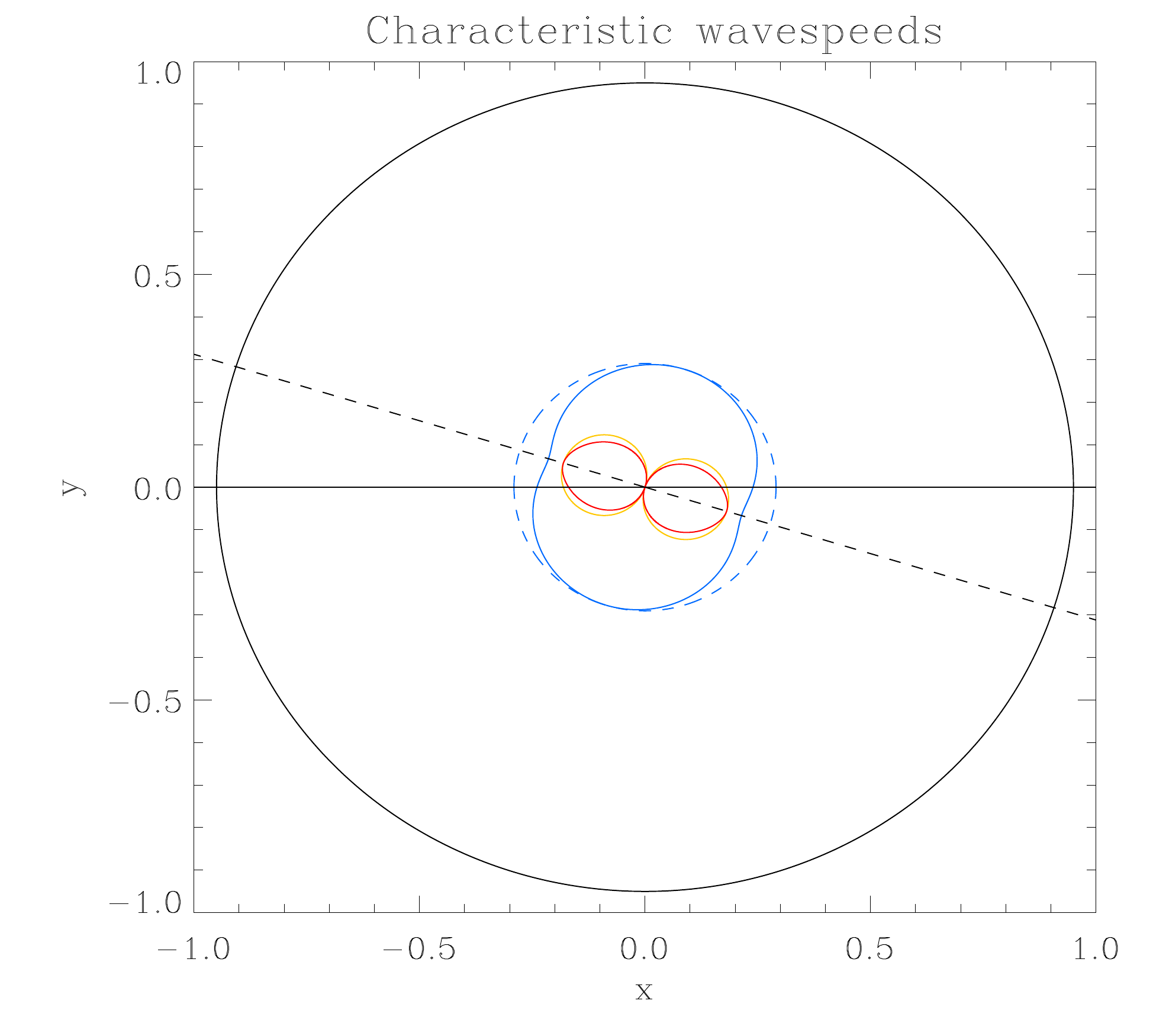}
\caption{Left panel: Characteristic wavespeed surfaces
of fast magnetosonic waves (in blue), { Alfv\'en} waves
(in yellow), slow magnetosonic waves (in red) and entropy waves (the
point at the origin) for the homogeneous state of a magnetized ideal
gas in the fluid rest frame corresponding to models PH02. The jet axis
is along the $x$ axis, whereas the oblique discontinuous straight line
is along the average helical magnetic field (in the fluid rest
frame). The black circumference represents the flow propagation
velocity. Right panel: The same as the left panel but for the
homogeneous state corresponding to KH10.}
\label{f:charwaves}
\end{minipage}
\end{center}
\end{figure*}

  The characteristic wave speed diagrams \citep{AM10}, or
phase polar diagrams \citep{CM15}, show
the normal speed of planar wave fronts propagating in different
directions, the speed given by the distance between the origin and
the normal speed surface along the corresponding direction. These
speeds correspond to the phase speeds of linear
perturbations studied by \cite{KM08}. The two panels of
Fig.~\ref{f:charwaves} display the characteristic wavespeed surfaces
of fast magnetosonic waves ($\lambda_F$, in blue), Alfv\'en waves
($\lambda_A$, in yellow), slow magnetosonic waves ($\lambda_S$, in
red) and entropy waves (the point at the origin) for the homogeneous
states of a magnetized ideal gas in the fluid rest frame corresponding
to models PH02 and KH10. The jet axis is along the $x$ axis, whereas
the oblique discontinuous straight line is along the average helical magnetic
field (in the fluid rest frame).

  Distinctly to sound waves, magnetosonic waves are anisotropic, with
the propagation speeds $\lambda_S$, $\lambda_F$ and $\lambda_A$
depending on the angle, $\chi$, of the propagation direction and the
magnetic field 

\begin{equation}
\lambda_A (\chi) = c_A \cos\chi
\end{equation}

\begin{equation}
\lambda_{F,S}(\chi) = \pm \sqrt{\frac{1}{2}\left[ d(\chi)^2 \pm (d(\chi)^4 - 4 c_s^2c_A^2
    \cos^2\chi)^{1/2} \right]}
\end{equation}
In these expressions, $d(\chi)^2 = c_s^2 + c_A^2 - c_s^2c_A^2 \sin^2
\chi$, where $c_s$ { ($ = \sqrt{\gamma p/\rho h}$, for an ideal gas
  with adiabatic index $\gamma$)} is the sound speed, and $c_A$ the Alfv\'en speed
\begin{equation}
c_A = \sqrt{\frac{b^2}{\rho h + b^2}}.
\label{e:cA}
\end{equation}

  An homogeneous flow is said to be super-fast magnetosonic if the
flow velocity, $v$, is $v > \lambda_F(\chi)$, where $\chi$ is the
angle between the flow propagation direction and the magnetic field in
the fluid rest frame. For practical purposes, we shall define our
superfast-magnetosonic jets as those for which
$v_j > c_{ms, \, j}$, with $c_{ms}$, the magnetosonic speed
(discontinuous blue line in Fig. \ref{f:charwaves}), being
\begin{equation}
c_{ms} = \sqrt{c_A^2 + c_s^2(1-c_A^2)}.
\label{e:cms}
\end{equation}
It can be seen that $c_{ms} \geq \lambda_F(\chi)$, $\forall \chi$.
In low magnetization jets, $\displaystyle{\beta = \frac{b^2}{2p}}$ is 
very small, which means that $c_A \ll c_s$ (see Eq.~\ref{e:cA}) and
$c_{ms} \approx c_s$ (see Eq.~\ref{e:cms}).

  Associated with the the superfast-magnetosonic flow is the
magnetosonic Mach number (see Cohen et al. 2015) 
\begin{equation}
{\cal M}_{ms} = \frac{W}{W_{ms}}\frac{v}{c_{ms}},
\end{equation}
where $W$ is the flow Lorentz factor and $W_{ms}$ is the flow Lorentz
factor associated to the magnetosonic speed. In low magnetization
jets, the magnetosonic Mach number coincides with the common (sound)
Mach number.

%
%

\end{document}